\documentclass[twocolumn,aps,prb,epsf,showpacs]{revtex4}
\usepackage{amsmath}
\usepackage{epsfig}
\usepackage{epsf}
\usepackage{array}
\usepackage{color}

\begin{document}
\bibliographystyle{apsrev}

\title{Charge density distribution and optical response of the  LaAlO$_{3}$/SrTiO$_{3}$ interface }
\author{Se Young Park}
\author{Andrew J. Millis}
\affiliation{Department of Physics, Columbia University, New York, NY 10027, USA}
\date{\today }

\begin{abstract}
We present calculations of the charge density profile, subband occupancy  and ellipsometry spectra of the electron gas at the LaAlO$_{3}$/SrTiO$_{3}$ interface. The calculations employ self-consistent Hartree and random phase approximations, a tight binding parametrization of the band structure and a model for the optical phonon of SrTiO$_3$.  The dependence of the spatial structure and occupancy of subbands on  the magnitude of the polarization charge at the interface and the dielectric function is determined. The interface-confined subbands may be labelled by the symmetry (xy, xz, or yz) of the $Ti$ d-orbitals from which they mainly derive.  The xy-derived band nearest the interface contains the major proportion of the electronic charge, but a large number of more distant, slightly occupied xy-derived bands are also found. Depending on the magnitude of the polarization charge, zero, one, or two xz/yz derived subbands are found. When present, these xz/yz bands give the dominant contribution to the long-distance tail of the interface charge.  The response to applied ac electric fields polarized parallel and perpendicular to the interface is calculated and the results are presented in terms of  ellipsometry angles. Two features are found: a dip in the spectrum near the LO feature of the STO phonon and a peak at the higher energy.  We show that the form and magnitude of the dip is related to the  Drude response of carriers moving in the plane of the interface while the peak arises from the plasmon excitation of the xz and yz electrons. The relation of the features of the subband occupancies and the in-plane conductivities is given.  
\end{abstract}


\maketitle

\section{Introduction}
Transition metal oxide materials have been investigated extensively because of their remarkable electronic properties, including  high Curie temperature magnetism, metal insulator transitions, spin and orbital order and superconductivity \cite{Imada98,Nagaosa00}.  In bulk materials these phases can be controlled by doping,  pressure and chemical composition  but the degree to which material properties can be varied is limited.  Progress over the last decade in  pulsed laser deposition and molecular bean epitaxy has provided new ways to control the properties of novel materials, by enabling the formation of atomic precision superlattices and interfaces unobtainable with bulk crystal growth methods \cite{Mannhart08,Zubko11,Hwang12}. Artificially created oxide systems  have been reported to exhibit a range of striking behavior including metal-insulator transitions \cite{thiel,Caviglia08,Perucchi10,Yoshimatsu10,Son10,Jang11,Moetakef11}, superconductivity \cite{Reyren07,Biscaras10}, and magnetism \cite{Brinkman07,Luders09,Ariando11,bert,Li11}.  The very short length scales and low volume carrier densities in these systems mean that characterization--of crystal structure, of defects, of electronic structure and of carrier density- is a key issue. 

Perhaps the most widely studied class of artifically created strongly correlated systems is the ``oxide interface'',\cite{Mannhart08,Zubko11} where in appropriate circumstances the junction between two insulating materials can support a two dimensional electron gas with interesting properties including superconductivity \cite{Reyren07,Biscaras10} and magnetism \cite{Brinkman07,Ariando11,Bark11,bert,Li11}. The most widely studied system is the  interface between LaAlO$_{3}$ (LAO) and SrTiO$_{3}$ (STO) \cite{ohtomo}, but variants in which the LAO is replaced by LaTiO$_3$ \cite{Ohtomo02,Biscaras10}, GdTiO$_3$ \cite{Moetakef11} or vacuum \cite{syro} or the STO by KTaO$_3$ \cite{King12} have also been studied.  A wide variety of experiments and theoretical calculations suggest that the electron gas  is confined to within $\sim10\; nm$ or less of the interface \cite{Okamoto06,Bristowe09,Pentcheva10,Biscaras10, dubroka}, resides in the STO in subbands derived  from the $Ti$ $t_{2g}$ symmetry d-orbitals \cite{Okamoto06,Bristowe09,syro,chen,son} and  is induced by the polar catastrophe scenario \cite{thiel,chen,son,Bristowe09,Pentcheva10,dubroka,copie}. However, the details of the value and spatial extent of the charge density as well as the subband structure and occupancy are not clearly understood. 

In the LAO/STO system the carrier density inferred from Hall effect and other transport measurements is typically one order of magnitude less than the number theoretically  expected from polar catastrophe \cite{thiel}, although high energy photoemisson measurements yield numbers more nearly consistent with theory \cite{Sing09}. Other contributions to the charge density besides the polar catastrophe mechanism have been proposed \cite{Siemons07,Chambers10} and a key role for defects has been documented \cite{nakagawa}. 

Inferences from the magnetic field dependence of the superconducting properties \cite{Caviglia08,Biscaras10} indicate that the spatial extent is $\sim 10-20nm$ whereas band theory calculations \cite{Okamoto06,Bristowe09,Pentcheva10} place most of the electrons substantially closer to the interface. Finally, the  $t_{2g}$ orbitals into which the electrons are believed to go have a substantial anisotropy in their dispersion, with one of the three bands dispersing much less strongly in the direction perpendicular to the interface than the others. The relative occupancy of the different orbitals therefore plays a crucial role in the spatial structure of the electron gas.

In this paper, we employ a minimal tight binding model   to describe the subband structure and occupancy of electrons at a polar interface. Our model includes charge self-consistency at the Hartree (self-consistent screening) level and the spatial and frequency structure of the dielectric constant of the host material as well as the orbital degeneracy of the $t_{2g}$ levels that host the relevant electrons. The subband structure and occupancies are computed as a function of the magnitude of the polarization charge at the interface. We find that most of the added charge goes into bands derived from $d_{xy}$ orbitals which lie very close to the interface, but that an important contribution to the density of charges far from the interface comes from electrons in the lowest-lying $d_{xz/yz}$-derived bands and that the form of the long distance charge distribution reflects the spatial structure of the lowest  subband wave function.  The dielectric tensor is then calculated within the random phase approximation (RPA) and is used in a solution of the relevant Maxwell equations to obtain expressions for the response to incident radiation polarized parallel and perpendicular to the interface. The results are expressed in terms of reflectivity spectra and ellipsometry angles.  The relation of the ellipsometry angles to the charge density, subband occupancies, and in-plane conductivity is determined.

The rest of this  paper is organized as follows. Section \ref{Model} introduces the tight binding/Hartree model and presents the theoretical formalism needed for the calculation of the dielectric tensor and reflectivity. Section \ref{SCHartree} presents the subband structure, charge density, and orbital occupancy following from the solution of our equations in different physically relevant cases.  Section \ref{Ellipsometry} is devoted to the analysis of the ellipsometry angles. Section \ref{Conclusion} is a summary and conclusion. 
	
\section{Formalism \label{Model}}
\subsection{Tight-binding model and self-consistent Hartree approximation}
In this subsection, we introduce a tight-binding model describing the electronic structure of electrons at an oxide interface. While our specific motivation is the electronic properties of the LAO/STO interface, the basic physics we address is that of electrons in $t_{2g}$-derived bands of a transition metal oxide material confined by a  Coulomb potential arising from a charge sheet. We therefore believe our results apply also to the STO/vacuum \cite{syro},  GTO/STO \cite{Moetakef11} and (with appropriate modifications) to the LAO/KTaO \cite{King12} systems. Extension to systems with $e_g$ derived bands would be straightforward but is not considered here. 

Bulk LAO and STO are  insulators with band gaps of about 5.6 eV and 3.2 eV, respectively\cite{lim, benthem}. They crystallize in forms which are small distortions of  the standard $ABO_3$ perovskite structure and we use the simple perovskite notation with axes $(001),~(010),~(100)$ aligned along the $B-O$ bonds  to indicate directions. We consider the most common experimental situation, in which an $(001)$ interface of a very large STO crystal is capped with some number $M$ of layers of LaAlO$_3$ in the same crystallographic orientation. When viewed along the $(001)$ direction, LAO is polar and if the interface is such that the TiO$_2$ plane of the STO adjoins the LaO plane of the LAO then basic electrostatic considerations \cite{Mannhart08,nakagawa,Schlom10} suggest that in the idealized situation of perfectly aligned defect-free materials, the electric fields caused by the polar character of the LAO induce an interface electron gas \cite{ohtomo,Mannhart08,nakagawa}. 

The areal charge density of the induced electron gas vanishes if the number of capping layers $M$ is less than a critical value of order $2-4$ and for $M$ greater than this critical value the areal charge density increases, eventually asymptotic to the value $0.5$ electrons per in-plane unit cell. Theory \cite{Bristowe09,Schlom10} and experiment \cite{nakagawa}  indicate that  the critical $M$ at which charge first appears, the thickness beyond which the induced charge becomes  close to the theoretical asymptotic value and indeed the asymptotic value itself  depend on the dielectric properties of host (STO) and overlayer (LAO) material, on the chemistry of the free surface of the LAO (which controls the energetics of the compensating charges) and on density of charged defects such as $O$ vacancies or interstitials.  For our purposes it is enough to model the polar interface as a semi-infinite block of STO, terminated in the (001) direction by a sheet of positive areal charge density. The magnitude $en_0/d^{2}$ of the areal charge density ($d^{2}$ is the area of a unit cell.) is a parameter of the model, and we consider a range of values between $n_0=0$ and $n_0=0.5$.

In the polar catastrophe scenario, for thick enough polar over layers, a charge of $n_{0}$ is transferred from the STO region to the outer boundary of the polar overlayer. The compensating electronic charge density $-n_{0}$ is trapped near the interface by a potential which has a contribution $V=4 \pi e^2n_0z/\epsilon d^{2}$ from the charge at the outer boundary of the polar over layer. This potential is reduced by the scale dependent dielectric parameter of STO and by screening from the interface electrons. In this situation the in-plane crystal momentum is a good quantum number while an orbital-dependent subband structure describes the motion perpendicular to the plane.  We treat this situation using a self-consistent Hartree approximation.  The band alignments of LAO and STO are such that it is energetically preferable to accommodate the trapped electrons in the STO \cite{Mannhart08,nakagawa,Bristowe09,Pentcheva10}, in bands derived from the $t_{2g}$ ($d_{xy,yz,xz}$) orbitals of the $Ti$. We model these bands in terms of a tight binding model with three orbitals per unit cell and only nearest neighbor hopping.  The nearest neighbor-only hopping assumption implies that the orbitals are not mixed on the single-particle level, and moreover implies that electrons in the $xy$-derived bands move only in the $xy$ plane etc. This drastically simplifies the ensuing theoretical treatment.  In the actual band structure, small farther neighbor hopping terms exist which mix the orbitals. However, the effects of these terms are small and do not affect our main conclusions, which have to do with charge distributions and average optical responses, in any important way.

With these considerations, we write the effective Hamiltonian of the system as 
\begin{eqnarray}
\hat H(\mathbf{k}) &=& \sum_{\substack{ \mathbf{k} \; i j  \\l=\{xy,yz,xz \}}} 
	t^{l}_{ij} (\mathbf{k}) \;C^{l\dagger}_{\mathbf{k} i} C^{l}_{\mathbf{k} j}  \nonumber \\
&& + \sum_{\substack{\mathbf{k}\;i \\ l=\{xy,yz,xz \}}} 
		(V^{H}_{i}[n] + V^{Lat}_{i})\;C^{l\dagger}_{\mathbf{k} i} C^{l}_{\mathbf{k} i} ~, 
\label{eq:H}
\end{eqnarray}
where $\mathbf{k}$ is the in-plane momentum, $i$ and $j$ are plane indices, $l$ denotes orbital indices, $t^{l}_{ij}(\mathbf{k})$ represents the  hopping between Ti atoms (orbitally diagonal, within our assumptions), $V^{H}_{i}[n]$ is  the self-consistent Hartree potential arising from the electron density $n$, and $V^{Lat}_{i}$ represents the external potential from the polarization charge. 

Because we are interested in low electron densities per orbital we approximate the hopping matrices $t^{l}_{ij}(\mathbf{k})$ as 
\begin{eqnarray}
t^{xy}_{ij}(\mathbf{k}) &=&td^{2} \vert \mathbf{k}\vert^{2} \delta_{ij}	\nonumber \\
t^{yz}_{ij}(\mathbf{k}) &=& td^{2} k^{2}_{y}\delta_{ij}
	- t(\delta_{ij-1} -2\delta_{ij}+ \delta_{ij+1})~, \nonumber \\
t^{xz}_{ij}(\mathbf{k}) &=&td^{2}k_{x}^{2}  \delta_{ij}
	- t(\delta_{ij-1} -2\delta_{ij}+ \delta_{ij+1})~,
\end{eqnarray}
Here we have made the small $k$ approximation to the dispersion and observe that corrections to the orbital-diagonal nearest-neighbor-only hopping approximation appear only to higher order in $k$.   $d$ is the distance between two $Ti$ atoms (any difference between the in-plane and transverse Ti-Ti distance is irrelevant for our purposes and is neglected). We set $t=0.34 \; eV$.  

The spatial dependence of the potential arising from the polar charge and the self-consistent field of the electrons is complicated by the large and strongly scale-dependent behavior of the dielectric function of STO\cite{hemberger, christen}. Within our tight binding approach we associate electrons with $Ti$ sites and the dielectric function becomes a link variable, depending on the distance of a given $Ti-Ti$ bond from the interface. The potential arising from the polarization charge then may be written as \begin{equation}
V_{i}^{Lat} 
= \frac{4\pi e^{2} }{d} n_{0}\sum_{j=1}^{i-1}\frac{1}{\epsilon_{jj+1}}~, 
\label{eq:Vlat}
\end{equation}
with $\epsilon_{jj+1}$ the dielectric constant appropriate to the region between the $Ti$-site in plane $j$ and the $Ti$ site in plane $j+1$.    The values $\epsilon_{jj+1}$ of the dielectric function are considered as parameters that are fitted by comparing the resulting density with existing DFT results\cite{chen}.  

Similarly, the Hartree potential $V^{H,l}_{i}$ is written as 
\begin{equation}
V_{i}^{H}[n] = -\frac{4\pi e^{2} }{d} \sum_{j=1}^{i-1} n_{j} \sum_{m=j}^{i-1}\frac{1}{\epsilon_{mm+1}}~,
\label{eq:VH}
\end{equation}
where $n_{j} $ represents the charge per unit cell in the j$^{th}$ TiO$_{2}$ plane, given as $ n_{j} = \langle \sum_{l}n^{l}_{\mathbf{k}=0, j} \rangle = \langle \sum_{\mathbf{q} l} C_{\mathbf{q}j}^{l\dagger}C^{l}_{\mathbf{q}j} \rangle$.  Charge neutrality implies $n_{0} = \sum_{i} n_{i}$. We note that the Hartree potential has no dependence on in-plane momentum in this approximation. . 

The charge densities are obtained from the solution of the  self-consistent Hartree equations; this is greatly simplified by the orbital-diagonal and nearest-neighbor-only  nature of the Hamiltonian.  The basic equations are
\begin{equation}
\sum_{j}\left[ t_{ij}^{l}(\mathbf{k}) + \delta_{ij}(V^{H}_{i}[n] + V^{Lat}_{i})  \right] \psi_{n j}^{l}(\mathbf{k})
= \varepsilon^{l}_{\mathbf{k} n} \psi_{n i}^{l}(\mathbf{k})~,
\label{eq:SCH}
\end{equation}
for  $\psi^{l}_{nj}(\mathbf{k})$ and $\varepsilon^{l}_{\mathbf{k}n}$, the wavefunction of the  nth eigenstate of  orbital $l$ and  the  corresponding eigenvalue, and 
\begin{equation}
n_{i} = \sum_{\substack{n \mathbf{q} \\ l=\{xy,yz,xz\}}} f_{F}(\varepsilon^{l}_{\mathbf{q}n})
	\vert \psi_{n i}^{l}(\mathbf{q}) \vert^{2}
\end{equation}
with the Fermi-Dirac distribution function, $f_{F}(\varepsilon^{l}_{\mathbf{q}n})$ and chemical potential chosen so that the total electronic charge per in-plane unit cell equals the polar charge $n_0$. 

Three features of the equations warrant comment. First, the spectrum involves both bound (to the interface) and unbound eigenfunctions. Second,  because in our formalism the $xy$ orbital disperses only in the plane parallel to the interface  the $n^{th}$ $xy$ band involves carriers in plane $n$ (measured from the interface) and has minimum energy $\varepsilon^{xy}_{\mathbf{k}=0n}=V^{Lat}_n+V^H_n$. At long distances from the interface the binding energy tends exponentially to zero with Thomas-Fermi decay length set by the long-distance behavior of the dielectric constant and by the hopping $t$. The very large value of the STO dielectric constant means that this length is in practice very long. Third, the $xz/yz$ eigenfunctions are delocalized over many planes, and as we shall see in reasonable cases only one bound-state subband is occupied.

\subsection{Current correlation function and dielectric tensor}
In this subsection, we calculate the current correlation function using self-consistent linear response (RPA) theory applied to the Hamiltonian of the previous section. We relate the obtained response functions to the transverse dielectric tensor and use them to calculate the reflectivity of the electron gas at the interface. In deriving the current correlation function, we assume that the impurity scattering may be treated within the  relaxation time approximation, i.e. may be modeled by a frequency and momentum-independent scattering time $\tau$\cite{Mermin70}. The resulting formula gives valid results both in diffusive ($\omega \ll 1/\tau$) and collision-less ($\omega \gg 1/\tau$) regimes.

The response functions depend on the polarization of the incident light, in other words on the direction of the applied electric field with respect to the plane of the interface. For in-plane electric fields, screening effects are negligible as are interband transition matrix elements (the small further-neighbor hopping terms neglected in our theory would give weak and unimportant interband terms). The response is therefore a standard Drude response in which each occupied subband contributes in parallel. Because the $yz$ ($xz$) band does not disperse in the $x$ ($y$) direction, the $x$-direction ($y$- direction) current has contributions only from the $xy$ and $xz$ ($xy$ and $yz$) orbitals. Therefore for the in-plane response we can expect that we can obtain information concerning the orbitally resolved electron density per in-plane unit cell, $n_{xy}+n_{xz}$ or $n_{xy}+n_{yz}$, as well as the in-plane conductivities.

The electronic response from out-of-plane electric fields arises only from the $xz/yz$ bands because in our approximation the $xy$ orbitals have only in-plane dispersion. Unlike the case of in-plane electric field, interband transitions (principally  to unbound  states) are important. In the physically relevant situation of an effectively infinitely thick STO substrate, the unbound states form a continuum. In our actual calculations we use a finite system with a large number $N$ (typically 40-80)  of STO layers but the modest impurity-induced broadening that we include means that our calculated results are a continuum whose form has negligible dependence on $N$ and broadening.

Computations of the optical response require the  gauge invariant current operator, which is obtained by inserting the vector potential $A$ into the Hamiltonian in the usual minimal coupling way and expanding to zeroth and first order in $A$ leading to  ``paramagnetic'' and ``diamagnetic'' terms which are expressed   in second quantized notation as
 \begin{eqnarray}
\mathbf{J}_{i}^{\parallel,p}(\mathbf{q}) &=& 2etd \sum_{\mathbf{p}mm^{\prime}}
	2(\mathbf{p}-\frac{\mathbf{q}}{2})F^{i}_{m^{\prime}m} 
	c^{\dagger}_{\mathbf{p}-\mathbf{q}m^{\prime}}c_{\mathbf{p}m}~,
	\nonumber \\
J_{i+\frac{1}{2}}^{\perp, p}(\mathbf{q}) &=& 2etd \sum_{\mathbf{p}mm^{\prime}}
	2Q^{i+\frac{1}{2}}_{m^{\prime}m}
	c^{\dagger}_{\mathbf{p}-\mathbf{q}m^{\prime}}c_{\mathbf{p}m} ~,\nonumber \\
\mathbf{J}^{\parallel, dia}_{i}(\mathbf{q})&=&- 2\frac{2e^{2}  t}{cd} \sum_{\mathbf{k} mm^{\prime}} 
	c^{\dagger}_{\mathbf{k}m^{\prime}}c_{\mathbf{k}m} 
	F^{i}_{m^{\prime}m} \mathbf{A}_{i}^{\parallel}(\mathbf{q},\omega) \nonumber \\
J^{\perp, dia}_{i+\frac{1}{2}  }(\mathbf{q}) &=& -2\frac{e^{2}t}{cd} \sum_{mm^{\prime}\mathbf{k}}
	c^{\dagger}_{\mathbf{k} m^{\prime}}c_{\mathbf{k}m}(\psi^{\ast}_{m^{\prime}i+1}\psi_{m i} \nonumber \\
&& +\psi^{\ast}_{m^{\prime}i}\psi_{m i+1})
	A^{\perp}_{i+\frac{1}{2}}(\mathbf{q})~,
\label{eq:def-j}
\end{eqnarray}
where the factor 2 in each equation is from the spin degeneracy, $c_{\mathbf{q}m}$ is an annihilation operator for $m^{th}$ subband eigenstate with momentum $\mathbf{q}$ satisfying $C^{l}_{\mathbf{q}i} = \sum_{m \in l} c_{\mathbf{q}m} \psi_{m i}^{l}$, the $\mathbf{A}$ is the vector potential, $c$ is the speed of light, the $\parallel$ and $\perp$ denote the  in- and out of- plane directions and the superscripts $p$ and $dia$ represent the paramagnetic and diamagnetic contributions to the  current, respectively. Because the hopping and hence the current is orbital-diagonal we have not explicitly denoted the orbital index here; this information is included in the  eigenvalue index $m$ runs from 1 to $3N$  with $N$ the number of TiO$_{2}$ layers included in the calculation including the orbital index. The current in $z$ direction is properly defined as a link variable; we use $i+\frac{1}{2}$  to denote the link between the $i^{th}$ and $(i+1)^{st}$ plane.  The current depends upon the wave functions; this information is encoded in the functions 
\begin{eqnarray}
F^{i}_{m^{\prime}m} = \psi^{\ast}_{m^{\prime} i} \psi_{mi}~,
\end{eqnarray}
and
\begin{eqnarray}
Q^{i+\frac{1}{2}}_{m^{\prime}m} &=& \frac{1}{2d} ( \psi^{\ast}_{m^{\prime}i+1}\psi_{m i}-\psi^{\ast}_{m^{\prime}i}\psi_{m i+1} ) \nonumber \\
&=& \frac{i(\varepsilon_{\mathbf{q}m^{\prime}}-\varepsilon_{\mathbf{q}m})}{2td}\sum_{j=1}^{i}F^{j}_{m^{\prime}m}~. 
\label{eq:def-Q}
\end{eqnarray}
To obtain the expression for $Q$ we have used  Eq.~(\ref{eq:SCH}). An important simplification following from our assumption of nearest-neighbor hopping is that both $F^{i}_{m^{\prime}m}$ and $Q^{i+\frac{1}{2} }_{m^{\prime}m}$ are independent of in-plane momentum $\mathbf{q}$ since eigenvectors $\psi_{mi}$ for eigenvalue $\varepsilon_{\mathbf{q}m}$ are the same for all value of $\mathbf{q}$ and the difference between two eigenvalues with the same $\mathbf{q}$ is independent of $\mathbf{q}$. 

We adopt the gauge in which the scalar potential vanishes and the vector potential is related to the electric field as $\mathbf{A} = \frac{c}{i\omega} \mathbf{E}$ and initially consider the clean limit in which the impurity scattering time $\tau \rightarrow \infty$ so that the problem is translation-invariant in the plane of the interface.  Linear response theory then yields the   paramagnetic and diamagnetic contributions to the gauge-invariant  conductivity tensor as 
\begin{eqnarray}
\hat \sigma_{ij}^{p} &=& 2\frac{ 4 i  e^{2} t^{2}d}{\omega N_{2D} } \sum_{\mathbf{k} m m^{\prime}}
	\frac{f_{F}(\varepsilon_{m^{\prime}\mathbf{k}}) - f_{F}(\varepsilon_{m\mathbf{k}+\mathbf{q}})  }{\omega + 
	\varepsilon_{m^{\prime} \mathbf{k} } - \varepsilon_{m\mathbf{k}+\mathbf{q}} + i\eta}\nonumber \\
&& \times
	\left[(\mathbf{k}+ \frac{\mathbf{q}}{2})F^{i}_{m^{\prime}m} + Q_{m^{\prime}m}^{i+\frac{1}{2}}\hat z \right] \nonumber \\
&& \times	
	\left[(\mathbf{k}+ \frac{\mathbf{q}}{2})F^{j}_{m m^{\prime}} + Q_{m m^{\prime}}^{j+\frac{1}{2}}\hat z \right]~, \nonumber \\
\hat \sigma^{dia}_{ij} &=& \frac{2ie^{2}t}{\omega d } \delta_{ij} 
	\Big[ 
	\sum_{m \in xy} n_{m}F^{i}_{mm} (\hat x \hat x + \hat y \hat y)  \nonumber \\
&+& \sum_{m \in xz}n_{m} F^{i}_{m m} \hat x \hat x 
	+ \sum_{m \in yz} n_{m}F^{i}_{mm} \hat y \hat y 
	\nonumber \\
&+&\frac{1}{2} \sum_{\substack{m\in xz\\ \text{or } yz} } n_{m} (\psi^{\ast}_{m i+1}\psi_{m i}+\psi^{\ast}_{m i}\psi_{m i+1}) \hat z \hat z
	\Big],
\label{eq:sigma-gen}
\end{eqnarray}
where the factor 2 in $\hat\sigma^{p}_{ij}$ is due to the spin degeneracy, $N_{2D}$ is number of unit cells in the plane, $\eta$ is an infinitesimally small positive number, and $n_{m}$ is defined as $\langle 2\sum_{\mathbf{q}} c^{\dagger}_{\mathbf{q}m} c_{\mathbf{q}m}\rangle$ including spin degeneracy. 

We note that in the limit of $q \rightarrow 0$, the off-diagonal tensor components such as $\hat x \hat y$ vanish since $\hat \sigma_{ij}^{p}(\mathbf{q},\omega) = -\hat \sigma_{ij}^{p}(\mathbf{-q},\omega)$. Moreover for $xy$ bands,  the clean limit assumption which will be relaxed later means that the conductivity comes only from diamagnetic term since $F^{i}_{m^{\prime}m} \propto \delta_{m^{\prime}m}$. In this limit the conductivity tensor is reduced to 
\begin{eqnarray}
&&\hat \sigma_{ij}^{p}(\omega) = \frac{4i e^{2}t^{2}d}{\omega \sqrt{N_{2D}}}
	\sum_{m \in xz \; k} \frac{2\Delta_{mm_{0}}f_{F}(\varepsilon_{k m_{0} })}{\omega^{2}-\Delta_{mm_{0}}^{2}+i\eta} \nonumber \\
&&\times\Big[
	2k^{2} F^{i}_{mm_{0}}F^{j}_{m_{0}m}
	\left( \hat x \hat x +  \hat y \hat y \right) +4Q_{mm_{0}}^{i+\frac{1}{2}} Q_{ m_{0}m}^{j+\frac{1}{2}}\hat z \hat z 
	 \Big] \nonumber \\
&& = \frac{4ie^{2}t^{2}dn_{m_{0}}}{\omega } \sum_{m \in xz} P_{m_{0}}^{m}(\omega) \nonumber \\
&&\times \bigg[ \frac{(k_{F}^{m_{0}})^{2}}{3}F^{i}_{mm_{0}}F^{j}_{m_{0}m}(\hat x \hat x + \hat y \hat y) 
	+2Q_{mm_{0}}^{i+\frac{1}{2}} Q_{ m_{0}m}^{j+\frac{1}{2}}\hat z \hat z  \bigg] \nonumber \\
&& \hat \sigma^{dia}_{ij}(\omega) = \frac{2ie^{2}t}{\omega d } \delta_{ij} n^{xy}_{i} (\hat x \hat x + \hat y \hat y) \nonumber \\
&& +\frac{2ie^{2}t}{\omega d} n_{m_{0}} \sum_{m \in xz} F^{i}_{m m_{0}}F^{j}_{m_{0}m}  (\hat x \hat x + \hat y \hat y) \nonumber \\
&& + \frac{16ie^{2 }t^{2}d n_{m_{0}}}{\omega}\sum_{m \in xz}\frac{Q^{i+\frac{1}{2}}_{m m_{0}} Q^{j+\frac{1}{2}}_{m_{0}m}   }{\Delta_{mm_{0}}} \hat z \hat z~,
\label{eq:sigma}
\end{eqnarray}
where we use the symmetry between xz and yz orbital,  take  the index $m$ to run over the eigenstates of xz orbital and assume that only the lowest yz and xz band, denoted as $m_{0}$, is occupied (this will be discussed in more detail below). Therefore $n_{m_{0}}$ becomes the number of electrons in $yz$ (or $xz$) bands per in-plane unit cell ($n_{m_{0}} = n_{yz} = n_{xz}$). In Eq.~(\ref{eq:sigma}) we define $n^{xy}_{i} = \sum_{m \in xy} n_{m} F^{i}_{mm}$ which is the number of electrons per in-plane unit cell in $xy$ band located at $i^{th}$ layer and 
\begin{eqnarray}
P^{m}_{m_{0}}(\omega) =  \frac{2\Delta_{mm_{0}}}{\omega^{2}-\Delta_{mm_{0}}^{2}+i\eta}~,
\end{eqnarray}
and have used the identities 
\begin{eqnarray}
&& \delta_{ij} F^{i}_{m_{0}m_{0}} = \sum_{m} F^{i}_{mm_{0}} F^{j}_{m_{0}m}~, \nonumber \\
&& \delta_{ij} (\psi^{\ast}_{m_{0} i+1} \psi_{m_{0} i} + \psi^{\ast}_{m_{0} i} \psi_{m_{0} i+1})  \nonumber \\
&&= 8td^{2}\sum_{m \in xz}\frac{Q^{i+\frac{1}{2}}_{m m_{0}} Q^{j+\frac{1}{2}}_{m_{0}m}  }{\Delta_{mm_{0}}} 
\end{eqnarray}
exploiting  $\psi_{m i}^{\ast} = \psi_{mi} $.
Combining paramagnetic and diamagnetic terms, we obtain  the conductivity in the limit  $q \rightarrow 0$ as 
\begin{eqnarray}
&& \hat \sigma_{ij}(\omega) =  \frac{2ie^{2}t}{\omega d} \bigg[\delta_{ij} n^{xy}_{i} (\hat x \hat x + \hat y \hat y) \nonumber \\
&&+ n_{m_{0}}\sum_{m \in xz }\left( \frac{2}{3} \varepsilon^{m_{0}}_{F}P^{m}_{m_{0}}(\omega)  + 1 
	\right) F^{i}_{mm_{0}}F^{j}_{m_{0}m}(\hat x \hat x + \hat y \hat y) \nonumber \\
&& + n_{m_{0}} \sum_{m \in xz} 
	\frac{8t \omega^{2} d^{2}Q^{i+\frac{1}{2}}_{m m_{0}} Q^{j+\frac{1}{2}}_{m_{0}m}   }{\Delta_{mm_{0}}(\omega^{2}- \Delta_{mm_{0}}^{2} + i\eta)}\hat z \hat z\bigg]~. 
\end{eqnarray}
We note that the resulting conductivity can be separated into  ``local'' (i.e. layer-diagonal) (superscript L) and non-local (layer changing) (superscript NL) contributions as 
$\hat \sigma_{ij} = \hat \sigma^{L}_{i}\delta_{ij}+ \hat \sigma^{NL}_{ij}$ which are given as 
\begin{eqnarray}
\hat \sigma^{L}_{i} &=& \frac{2ie^{2}t}{\omega d} \left[ (n^{xy}_{i}+n^{xz}_{i})\hat x \hat x +(n^{xy}_{i}+n^{yz}_{i})\hat y \hat y)\right] \nonumber \\
&\equiv& \frac{\Omega^{\parallel 2}_{i}}{4\pi \omega} (\hat x \hat x + \hat y \hat y)~, \nonumber \\
\hat \sigma^{NL}_{ij} &=&\frac{2ie^{2}t}{\omega d} \bigg[
	\frac{2}{3}\epsilon^{m_{0}}_{F}n_{m_{0}}P^{m}_{m_{0}}(\omega) F^{i}_{mm_{0}}F^{j}_{m_{0}m}(\hat x \hat x + \hat y \hat y) \nonumber \\
&& +\frac{8n_{m_{0}}t \omega^{2} d^{2}Q^{i+\frac{1}{2}}_{m m_{0}} Q^{j+\frac{1}{2}}_{m_{0}m}   }{
	\Delta_{mm_{0}}(\omega^{2}- \Delta_{mm_{0}}^{2} + i\eta)}\hat z \hat z \bigg] \nonumber \\
&\equiv& \frac{\Omega_{\perp}^{2}}{4\pi \omega} 
	\frac{2\epsilon^{m_{0}}_{F}}{3}P^{m}_{m_{0}}(\omega) F^{i}_{mm_{0}}F^{j}_{m_{0}m}(\hat x \hat x + \hat y \hat y) \nonumber \\
&&+\frac{\Omega_{\perp}^{2}}{4\pi t} \omega P_{m_{0}}^{m}Z^{i+\frac{1}{2}}_{m m_{0}} Z^{j+\frac{1}{2}}_{m_{0}m}  \hat z \hat z~,
\end{eqnarray}
where we define $n^{xz}_{i} = n^{yz}_{i} = n_{m_{0}}F^{i}_{m_{0}m_{0}}$ and in-plane plasma frequency 
$\Omega^{\parallel 2}_{i} \equiv 8\pi e^{2}t (n^{xy}_{i}+n^{yz}_{i})/d$ for local conductivity, and introduce  $\Omega_{\perp}^{2} \equiv 8\pi e^{2}t n_{m_{0}}/d$ and $Z^{i+\frac{1}{2}}_{mm_{0}} = -\sum_{j=1}^{i} F^{j}_{mm_{0}}$ for non-local conductivity with the use of Eq.~(\ref{eq:def-Q}). For electric field independent of layer index (as is the case for radiation polarized parallel to the interface), the first term of $\hat \sigma^{NL}_{ij}$ is zero $(\sum_{j} F^{j}_{mm_{0}}=0)$ and thus for optical transitions, we can ignore the first term. Then the structure of the conductivity tensor becomes much simpler: the in-plane response is given by the sum of the Drude conductivities of each layer  and the out-of-plane response is from inter-subband transitions. The oscillator strengths in the conductivities are characterized by the  plasma frequencies $\Omega^{\parallel}_{i}$ and $\Omega_{\perp}$.  

We now include impurity scattering within the relaxation time approximation.\cite{Mermin70}  From the relation $\hat \epsilon_{ij}= \epsilon^{0} \mathbf{1}\delta_{ij} + \frac{4\pi i}{\omega } \hat \sigma_{ij}$ with $\epsilon^{0}$ defined as the bare dielectric constant without interface electrons, we obtain the transverse dielectric tensor with relaxation time $1/\tau$ as 
\begin{eqnarray}
\hat \epsilon_{ij}(\omega) &=& 
\epsilon^{0}(\omega)\delta_{ij} \mathbf{1} - \frac{(\Omega^{\parallel}_{i})^{2}}{\omega(\omega+i/\tau)} \delta_{ij} (\hat x \hat x + \hat y \hat  y) \nonumber \\
&& - \frac{\Omega_{\perp}^{2}}{t^{2}} \sum_{m \in xz } L_{m}(\omega)
	Z^{i+\frac{1}{2}}_{m m_{0}} Z^{j+\frac{1}{2}}_{m_{0}m}  \hat z \hat z~, 
\label{eq:ep}
\end{eqnarray}
where
\begin{eqnarray}
L_{m}(\omega) = \frac{2 t \Delta_{mm_{0}} }{\omega^{2}- \Delta_{mm_{0}}^{2}+ i\omega/\tau}~.
\end{eqnarray}
In the regime where $\omega \gg 1/\tau$, the dielectric response in the direction perpendicular to the plane is characterized by a series of inter-band peaks broadened by $\delta \omega \sim 1/2\tau$. On the other hand, for $\omega \ll 1/\tau$, we have diffusion poles where the imaginary part of dielectric function goes to zero in the limit of $\omega \rightarrow 0$.  In Eq.~(\ref{eq:ep}) we find that dielectric tensor in the direction perpendicular to the interface is finite in the limit of $\omega \rightarrow 0$, while  for in-plane direction it is proportional to $1/\omega$. This comes from the our assumption that  electrons at the interface are confined in the direction perpendicular to the interface but  may propagate freely in the plane of the interface. 

The calculations of the  reflectivity spectra in the next section require  the inverse of  $\epsilon^{zz}_{ij}$. From the expression in the Eq.~(\ref{eq:ep}), we can write the inverse of the dielectric tensor as 
\begin{eqnarray}
&&\epsilon^{zz -1}_{ij}\epsilon^{0} = \delta_{ij} + \tilde \Omega_{\perp}^{2}  \sum_{m \in xz }L_{m}
	Z^{i+\frac{1}{2}}_{m m_{0}} Z^{j+\frac{1}{2}}_{m_{0}m}   \nonumber \\
&&+ \tilde \Omega_{\perp}^{4} \sum_{\substack{m m^{\prime} \in xz \\ i_{1} }}L_{m}
	Z^{i+\frac{1}{2}}_{m m_{0}} Z^{i_{1}+\frac{1}{2}}_{m_{0}m}  L_{m^{\prime}}
	Z^{i_{1}+\frac{1}{2}}_{m^{\prime} m_{0}} Z^{j+\frac{1}{2}}_{m_{0}m^{\prime}}  \nonumber \\ 
&&+ \cdots \nonumber \\
&&= \delta_{ij} + \tilde \Omega_{\perp}^{2}  \sum_{mm^{\prime}\in xz} Z^{i+\frac{1}{2}}_{m m_{0}} L_{m}
	\Big[
	\delta_{mm^{\prime} }  + \tilde \Omega_{\perp}^{2}S_{mm^{\prime}}L_{m^{\prime}} \nonumber \\
&&+ \tilde \Omega_{\perp}^{4}\sum_{m_{1} \in xz }S_{mm_{1}}L_{m_{1}}S_{m_{1}m^{\prime}}L_{m^{\prime}} + \cdots
	\Big] Z^{j+\frac{1}{2}}_{m_{0}m^{\prime}} \nonumber \\
&&=  \delta_{ij} + \tilde \Omega_{\perp}^{2}  \sum_{mm^{\prime}\in xz} Z^{i+\frac{1}{2}}_{m m_{0}}  
	\Bigg[
	\hat L^{-1} - \tilde \Omega_{\perp}^{2}\hat S  
	\Bigg]^{-1}_{mm^{\prime}} Z^{j+\frac{1}{2}}_{m_{0}m^{\prime}} ~, \nonumber \\
\end{eqnarray}
where we define
\begin{eqnarray}
S_{mm^{\prime}} &\equiv& \sum_{i}Z^{i+\frac{1}{2}}_{m m_{0}} Z^{i+\frac{1}{2}}_{m_{0}m^{\prime}} \nonumber \\
\hat L \vert_{mm^{\prime}} &\equiv& L_{m} \delta_{mm^{\prime}} \nonumber \\
\tilde \Omega^{2}_{\perp} &\equiv& \frac{ \Omega^{2}_{\perp}}{\epsilon^{0}t^{2} }~.
\end{eqnarray}
We introduce the Coulomb matrix $v_{mm^{\prime}}(q)$, which in the limit of $q \rightarrow 0$ is related with $S_{mm^{\prime}}$ as: 
\begin{eqnarray}
\lim _{q \rightarrow 0}v_{mm\prime}(q) &=& -\lim_{q \rightarrow 0}
	\frac{2\pi e^{2}}{\epsilon_{0 }qd^{2}} \sum_{ij} F^{i}_{mm_{0}}e^{-qd\vert i-j\vert}  F^{j}_{m_{0}m^{\prime}} \nonumber \\
&=& \frac{4 \pi e^{2}}{\epsilon^{0}d} S_{mm^{\prime}}~.
\end{eqnarray}
With the definition $\epsilon^{0}  [\Delta\epsilon^{zz}]^{-1}_{ij} = \epsilon^{0}\epsilon^{zz -1}_{ij}- \delta_{ij}$ we have 
\begin{eqnarray}
(\epsilon^{0})^{2}  [\Delta\epsilon^{zz}]^{-1}_{ij}  = E_{C}^{0}\sum_{mm^{\prime}\in xz}  
	Z^{i+\frac{1}{2}}_{m m_{0}} \Pi_{mm^{\prime}} Z^{j+\frac{1}{2}}_{m_{0}m^{\prime}}~,
\label{eq:ZPiZ}
\end{eqnarray}
where
\begin{eqnarray}
\Pi_{mm^{\prime}} = 2n_{m_{0}}  \left[  t \hat L^{-1}- 2n_{m_{0}}\hat v \right]^{-1}_{mm^{\prime}}~,
\label{eq:PI-mm}
\end{eqnarray}
and $E_{C}^{0}$ is a bare charging energy defined as $4\pi e^{2}/d$.
In the following section we show that  $(\epsilon^{0}(\omega))^{2}[\Delta\epsilon^{zz}(\omega)]^{-1}_{ij}$ functions as an effective dielectric constant in the expressions for the reflectivity. We also note that the poles of density correlation function $\Pi_{mm^{\prime}}$ in Eq.~(\ref{eq:ZPiZ}) represent the inter-band plasmon excitations. These  are shifted from the inter-band transition energy due to the Coulomb interaction.  

If the  perturbing electric field is very smooth compared with the system size, it is reasonable to approximate 
$ [\Delta\epsilon^{zz}(\omega)]^{-1}_{ij}  \simeq \sum_{ij}  [\Delta\epsilon^{zz}(\omega)]^{-1}_{ij} \equiv \langle\langle \Delta\epsilon_{zz}^{-1}(\omega) \rangle\rangle$, implying 
\begin{eqnarray}
(\epsilon^{0})^{2}\langle\langle \Delta\epsilon_{zz}^{-1} \rangle\rangle =  E_{C}^{0} \sum_{mm^{\prime}\in xz}  \langle m_{0}\vert z \vert m \rangle  \Pi_{mm^{\prime}} \langle m^{\prime}\vert z \vert m_{0} \rangle~,
\label{eq:Pi-appr}
\end{eqnarray}
using $\sum_{i}Z^{i+\frac{1}{2} }_{mm_{0} } = \sum_{i} i F^{i}_{mm_{0}} = \langle m_{0} \vert z \vert m \rangle$ which is the dipole matrix element between the ground and excited states. Thus in the long-wavelength limit, we can expect that the structure of ($\epsilon^{0}(\omega))^{2}\langle\langle \Delta\epsilon_{zz}^{-1}(\omega) \rangle\rangle $ depends on plasmon poles and the dipole matrix element between occupied and excited states.

\begin{figure}[htbp]
\begin{center}
\includegraphics[width=0.9\columnwidth, angle=-0]{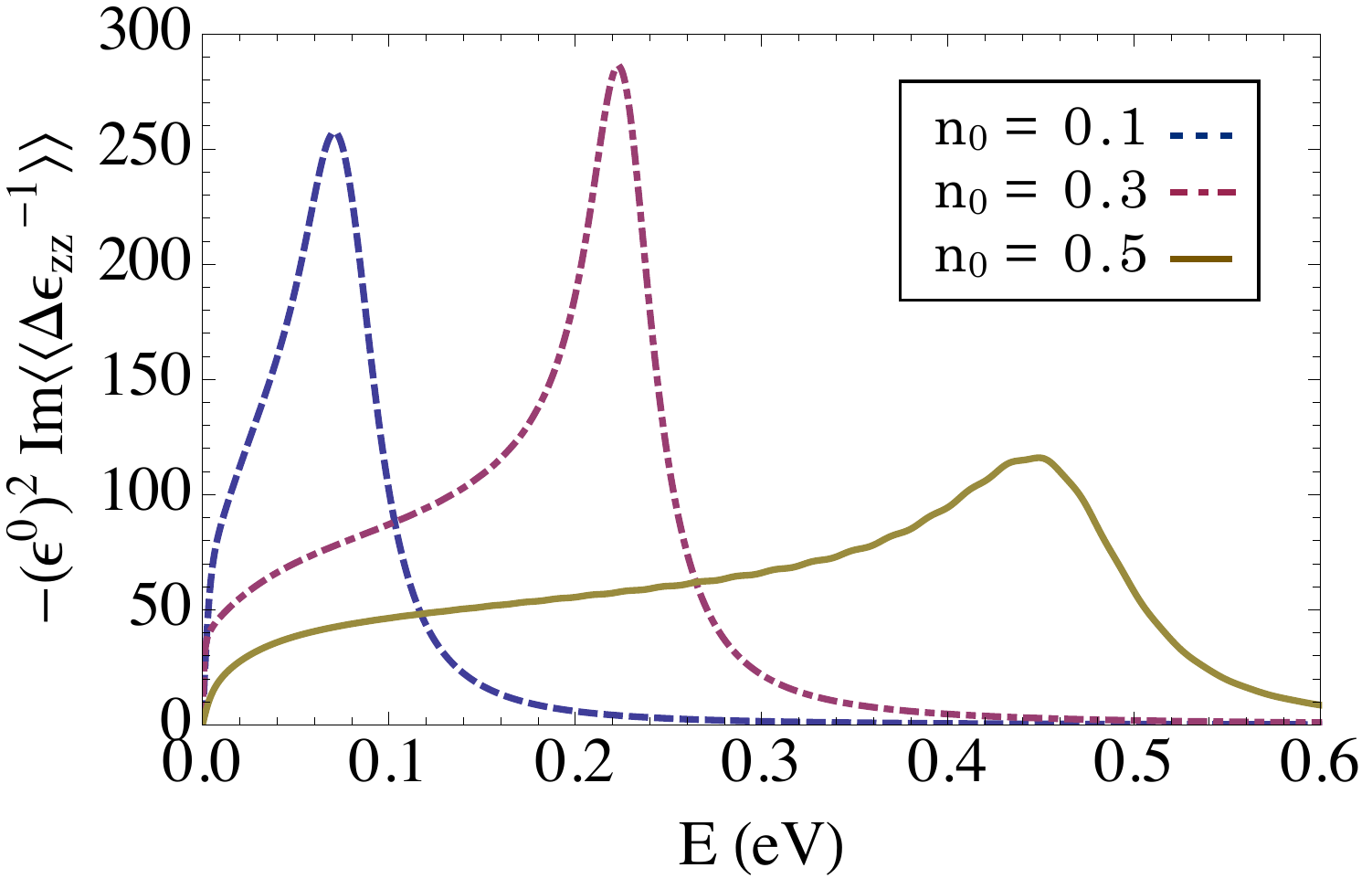}
\caption{(color online) Imaginary part of the  $(\epsilon^{0})^{2}\langle\langle \Delta\epsilon_{zz}^{-1}(\omega) \rangle\rangle $ for polar charge density of 0.1, 0.3, and 0.5 electrons per in-plane unit cell. For simplicity we set as a constant $\epsilon^{0} =  \lim_{\omega \rightarrow \infty} \epsilon_{S}^{0}(\omega)= 5.1$ where $\epsilon_{S}^{0}(\omega)$ is the dielectric constant of bulk STO. The scattering rate is set to $1/\tau = 40 \; meV$ for all three cases.}
\label{fig:0.5}
\end{center}
\end{figure}

In order to investigate the characteristics of plasmon excitation, we calculate $(\epsilon^{0})^{2}\langle\langle \Delta\epsilon_{zz}^{-1}\rangle\rangle$ from the solution of self-consistent Hartree equation (Eq.~(\ref{eq:SCH})) by assuming typical values for the static dielectric constant $\{ \epsilon_{jj+1} \}$ which will be discussed in Section \ref{SCHartree}. For simplicity we set $\epsilon^{0}(\omega)$ as a constant value ($\epsilon^{0} \equiv 5.1$) that is the high frequency limit of the dielectric constant of STO.
Fig.~\ref{fig:0.5} shows the imaginary part of the $(\epsilon^{0})^{2}\langle\langle \Delta\epsilon_{zz}^{-1} \rangle\rangle$ for polar charge density $n_{0}$ of 0.1, 0.3, and 0.5 electron per unit cell. We can see a well-defined peak for all three cases. The peak energy becomes higher and the peak width become broader as polar charge density increases. In following sections, we will show that this plasmon excitation produces high energy features in ellipsometry angles.

\subsection{Reflectance and ellipsometry spectra}
In this subsection, we obtain formulae for reflectivity  by solving Maxwell's equation using electromagnetic Green functions\cite{Dahl77,Bagchi79} and  the dielectric function obtained in the previous subsection. We consider two types of incident radiation: s-polarized, where  the incoming and outgoing electric fields are both parallel to the interface plane and the orthogonal  p-polarization in which we have both in-plane and out-of-plane components of electric fields. 

From the Maxwell equation, we can write the differential equation for the electric field and current density as 
\begin{eqnarray}
\nabla \times (\nabla \times \mathbf{E}) &=& -\frac{1}{c}  \frac{ \partial }{\partial t}\left[\frac{4\pi}{c}\mathbf{J} 
	+  \frac{1}{c}\frac{ \partial \mathbf{E}  }{\partial t}\right]~.
\end{eqnarray}
By introducing the dielectric tensor $\hat\epsilon(\omega,\mathbf{r},\mathbf{r}^{\prime}) 
= \hat \epsilon^{0}(\omega,z)\delta(z-z^{\prime}) +  \Delta \hat \epsilon(\mathbf{r},\mathbf{r}^{\prime})$ and transforming to frequency space, we can rewrite the equation as 
\begin{eqnarray}
&& \left[-\nabla \times \nabla\times + \frac{\omega^{2}}{c^{2}} \hat \epsilon^{0}(\omega, z) \right] \mathbf{E}(\omega,\mathbf{r}) \nonumber \\
&&= -\frac{\omega^{2}}{c^{2}}  \int d^{3}r^{\prime} \Delta \hat \epsilon(\omega,\mathbf{r},\mathbf{r}^{\prime}) \mathbf{E}(\omega,\mathbf{r}^{\prime})~. 
\label{eq:deq}
\end{eqnarray}
In Eq.~(\ref{eq:deq}), the contribution to the dielectric tensor of the  interface electrons is denoted as $\Delta\hat \epsilon (\omega,\mathbf{r}, \mathbf{r}^{\prime})$. It is in general non-local in space but vanishes in regions where the electron density vanishes.  We assume that the other contributions, such as phonon modes in STO, are local in space and are included in $\hat \epsilon^{0}$ and define the bare dielectric constant of LAO/STO heterostructures as $\epsilon^{0}(\omega,z) = \epsilon^{0}_{L}(\omega) \hat 1$ for $z < 0$ (i.e. in the LAO region) and  $\hat \epsilon^{0}(\omega, z) = \epsilon_{S}^{0}(\omega) \hat 1$ for $z>0$ (i.e. in the STO region).  
In addition,  we can write the electric field components as $\mathbf{E}(\omega,\mathbf{r})  = \mathbf{E}(\omega,z)e^{i\mathbf{Q} \cdot \boldsymbol \rho - i\omega t}$ with in-plane wave vector $\mathbf{Q}$ and position $\boldsymbol \rho$ where  $\vert \mathbf{Q} \vert = \sqrt{\epsilon^{0}_{L}}\omega \sin \theta /c $ with incident angle $\theta$. In the rest of this section we will omit the frequency dependence in the expressions for the electric fields and dielectric constants for simplicity. Solving the Eq.~(\ref{eq:deq}) is similar to solving a scattering problem where we have non-local dielectric function in the place of the scattering potential and we can write the general solution as\cite{Bagchi79} 
\begin{eqnarray}
\mathbf{E}(z) &=& \mathbf{E}^{0}(z) \nonumber \\
&-& \frac{\omega^{2}}{c^{2}}\int dz_{1} dz_{2} \; \hat G(z,z_{1}) \Delta \hat \epsilon(z_{1},z_{2}) \mathbf{E}(z_{2})~,
\label{eq:Escf}
\end{eqnarray}
where 
\begin{equation}
\left[-\nabla \times \nabla\times + \frac{\omega^{2}}{c^{2}} \hat \epsilon^{0}(z) \right] \hat G(z,z^{\prime}) = \hat{1}\delta(z-z^{\prime})~,
\label{eq:GFeq}
\end{equation}
and
$\mathbf{E}^{0}(z)$ satisfies
\begin{equation}
\left[-\nabla \times \nabla\times + \frac{\omega^{2}}{c^{2}} \hat \epsilon^{0}(z) \right] \mathbf{E}^{0}(z) = 0~.
\label{eq:E0}
\end{equation}
The explicit expression of $\hat G$ and $\mathbf{E}^{0}$ for $s$- and $p$-polarized lights are presented in appendix A.

For s-polarized light, the incoming and outgoing electric fields are both perpendicular to the plane of incidence and we take the direction of the electric field to be the  $y$ direction. Eq.~(\ref{eq:Escf}) then becomes 
\begin{eqnarray}
E_{y}(z) &=& E^{0}_{y}(z)  \nonumber \\
&-& \frac{\omega^{2}}{c^{2}}\int dz_{1} dz_{2}\; G_{yy}(z,z_{1})\Delta \epsilon_{yy}(z_{1},z_{2})E_{y}(z_{2})~.\nonumber \\
\label{eq:scf-spol}
\end{eqnarray}
Given the system size $L \sim 10nm$, the wave number times system size $kL$ is estimated as $\vert k L\vert \sim \vert 4\pi\sqrt{\epsilon^{0}} \alpha_{F} \frac{\hbar \omega}{   E_{c}^{0}}\frac{L}{d} \vert \ll 1 $, where $\alpha_{F}$ is the fine structure constant.  This means that neither the Green function nor the electric field changes significantly in the region where interfacial electrons reside. Thus, we can approximate the Eq.~(\ref{eq:scf-spol}) as 
\begin{eqnarray}
E_{y}(z) = E^{0}_{y}(z)-\frac{\omega^{2}}{c^{2}} G_{yy}(z,0)\langle \langle \Delta \epsilon_{yy}\rangle \rangle E_{y}(0)~,
\label{eq:Ey-appr-epyy}
\end{eqnarray}
where
\begin{eqnarray}
\langle \langle \Delta \epsilon_{yy}\rangle \rangle  &=& \int dz_{1}dz_{2} \; \Delta \epsilon_{yy}(z_{1},z_{2}) \nonumber \\
&=& - \frac{\Omega_{\parallel}^{2}}{\omega(\omega+i/\tau)}
\label{eq:scf-spolapp}
\end{eqnarray}
and $\Omega_{\parallel}^{2}$ is defined as  $ \sum_{i} \Omega^{\parallel 2}_{i}$ or $ 8\pi e^{2}t (n_{xy}+n_{yz})/d$.  We note that in the clean limit only diamagnetic terms contributes to the approximated dielectric tensor since the non-local terms give a contribution of the order of $(kL)^{2}$. Therefore, we expect a Drude response of the interface electrons for s-polarized light.  By substituting $z=0$ on both sides of Eq.~(\ref{eq:scf-spolapp}), we can obtain the $E_{y}(0)$ and write the electric field as 
\begin{eqnarray}
E_{y}(z) &=& e^{iqz} + r_{s}^{0} e^{-iqz} \nonumber \\
&-& \frac{\omega^{2}d(1+r_{s}^{0})}{2iq c^{2}}
	\frac{\langle \langle \Delta \epsilon_{yy}\rangle \rangle (1+r_{s}^{0})}{1+ \frac{(1+r_{s}^{0}) \omega^{2}d}{2iq c^{2}} 
	\langle \langle \Delta \epsilon_{yy}\rangle \rangle } e^{-iqz}  \nonumber \\
&=& e^{iqz} +r_{s}^{0}e^{-iqz}\bigg(1 + \nonumber \\
&&\frac{2iqd}{\epsilon^{0}_{L}-\epsilon^{0}_{S}} \frac{\langle \langle \Delta \epsilon_{yy}\rangle \rangle }{1+\frac{(1+r_{s}^{0}) qd }{2i\epsilon_{L}^{0}\cos^{2}\theta} \langle \langle \Delta \epsilon_{yy}\rangle \rangle }
\bigg)~,
\label{eq:Ey-loc}
\end{eqnarray}
where $r_{s}^{0}$ is the reflectivity without interfacial electrons which is $\frac{q-k}{q+k}$, and $q$ and $k$ are wave vectors in $z$-direction for  $z<0$ and $z>0$ defined as $\omega\sqrt{\epsilon_{L}^{0} \cos \theta} /c$ and $\omega \sqrt{\epsilon_{S}^{0}-\epsilon_{L}^{0}\sin\theta}/c$, respectively, for an incident angle $\theta$.  In Eq.~(\ref{eq:Ey-loc}), the relation $(1+r_{s}^{0} )^{2}= 4q^{2} r_{s}^{0}\frac{c^{2}/\omega^{2}}{(\epsilon^{0}_{L}-\epsilon^{0}_{S})} $ is used. From the definition of the reflectivity $E_{y}(z) = e^{iqz} + r_{s}e^{-iqz}$ we have  
\begin{eqnarray}
r_{s} = r_{s}^{0}\left(
1+\frac{2iqd}{\epsilon^{0}_{L}-\epsilon^{0}_{S}} \frac{\langle \langle \Delta \epsilon_{yy}\rangle \rangle }{1+\frac{(1+r_{s}^{0}) qd }{2i\epsilon_{L}^{0}\cos^{2}\theta} \langle \langle \Delta \epsilon_{yy}\rangle \rangle }
\right)~.
\label{eq:rs-appr}
\end{eqnarray}
We note that the reflectivity becomes $-1$ in the low frequency limit  and $r_{s}^{0}$ in the high frequency limit as expected. 

For $p$-polarized incident light, we have both in-plane and out-of-plane components of electric fields. Assuming that the electric field is in the plane of incidence ($xz$ plane), we can write\cite{Bagchi79} 
\begin{eqnarray}
E_{x}(z) &=& E^{0}_{x}(z) \nonumber \\
&-&\frac{\omega^{2}}{c^{2}}\int dz_{1} dz_{2} [ G_{xx}(z,z_{1})\Delta \epsilon_{xx}(z_{1},z_{2})E_{x}(z_{2}) \nonumber \\
&+& G_{xz}(z,z_{1})\Delta \epsilon_{zz}(z_{1},z_{2})E_{z}(z_{2})] \nonumber \\
E_{z}(z) &=& E^{0}_{z}(z) \nonumber \\
&-&\frac{\omega^{2}}{c^{2}}  \int dz_{1} dz_{2}[ G_{zx}(z,z_{1})\Delta \epsilon_{xx}(z_{1},z_{2})E_{x}(z_{2})\nonumber \\
&&+ G^{\prime}_{zz}(z,z_{1})\Delta \epsilon_{zz}(z_{1},z_{2})E_{z}(z_{2})] \nonumber \\
&-& \frac{1}{\epsilon^{0}(z)}  \int dz_{1} \Delta \epsilon_{zz}(z,z_{1})E_{z}(z_{1})~.
\label{eq:scf-ppol}
\end{eqnarray}
Using that the displacement field, defined as $D^{0}_{z}(z) = \epsilon^{0}(z)E^{0}_{z}(z)$, is continuous across the interface, we can simplify Eq.~(\ref{eq:scf-spol}) in terms of the displacement field as 
\begin{eqnarray}
E_{x}(z) &=& E^{0}_{x}(z) \nonumber \\
&-&\frac{\omega^{2}}{c^{2}}\int dz_{1} dz_{2} \Big[ G_{xx}(z,z_{1})\Delta \epsilon_{xx}(z_{1},z_{2})E_{x}(z_{2}) \nonumber \\
&&- G_{xz}(z,z_{1})\epsilon^{0}(z_{1})\Delta \epsilon^{-1}_{zz}(z_{1},z_{2})D_{z}(z_{2})\Big] \nonumber \\
D_{z}(z) &=& D^{0}_{z}(z) -\frac{\epsilon^{0}(z)\omega^{2}}{c^{2}}\int dz_{1} dz_{2} \nonumber \\
&& \times \bigg[ G_{zx}(z,z_{1})\Delta \epsilon_{xx}(z_{1},z_{2})E_{x}(z_{2}) \nonumber \\
&&- G^{\prime}_{zz}(z,z_{1})\epsilon^{0}(z_{1})\Delta \epsilon^{-1}_{zz}(z_{1},z_{2})D_{z}(z_{2})\bigg] ~,
\label{eq:scf-ppol-d}
\end{eqnarray}
where we have used 
\begin{eqnarray}
\epsilon^{0}(z)\Delta \epsilon^{-1}_{zz}(z,z_{1}) =-\int dz_{2} \Delta\epsilon_{zz}(z,z_{2})[\epsilon^{-1}]_{zz}(z_{2},z_{1})~. \nonumber \\
\end{eqnarray}
By applying the same approximation as in the $s$-polarization case (Eqs.~(\ref{eq:Ey-appr-epyy}) and (\ref{eq:scf-spolapp})), we obtain the coupled differential equations
\begin{eqnarray}
E_{x}(z) &=&  E_{x}^{0}(z) - \frac{\omega^{2}}{c^{2}} G_{xx}(z,0)\langle\langle \Delta\epsilon_{xx} \rangle\rangle E_{x}(0) \nonumber \\
&+& \frac{\omega^{2}}{c^{2}} G_{xz}(z,0) \epsilon^{0}_{S} \langle\langle \Delta\epsilon_{zz}^{-1} \rangle\rangle D_{z}(0) \nonumber \\
D_{z}(z) &=& D_{z}^{0}(z) - \frac{\omega^{2}}{c^{2}} \epsilon^{0}_{L} G_{zx}(z,0)\langle\langle \Delta\epsilon_{xx} \rangle\rangle E_{x}(0) \nonumber \\
&+& \frac{\omega^{2}}{c^{2}} \epsilon^{0}_{L}G_{zz}(z,0) \epsilon^{0}_{S} \langle\langle \Delta\epsilon_{zz}^{-1} \rangle\rangle D_{z}(0)~,
\label{eq:rp-loc}	
\end{eqnarray}
in which the dielectric properties of the materials are encoded in the two functions $\langle\langle\Delta\epsilon_{xx} \rangle\rangle$ (defined in Eq.~(\ref{eq:scf-spolapp})) which captures the in-plane Drude conductivity and $\langle\langle \Delta\epsilon_{zz}^{-1} \rangle\rangle$ (defined in Eq.~(\ref{eq:Pi-appr})) which represents the plasma excitation. Substituting $z=0$ leads to coupled equations for $E_{x}(0)$ and $D_{z}(0)$ that can be easily solved. By substituting the solutions back into Eq.~(\ref{eq:rp-loc}), we can get the reflectivity $r_{p}$ defined as
\begin{eqnarray}
E_{x}(z) = e^{iqz} -r_{p} e^{-iqz}~.
\end{eqnarray}
Then the reflectivity expressed in terms of averaged dielectric tensor $\langle\langle\Delta\epsilon_{xx} \rangle\rangle$  and $\langle\langle \Delta\epsilon_{zz}^{-1} \rangle\rangle$ is 
\begin{eqnarray}
 &r_{p}& = r_{p}^{0}+\frac{\frac{qd (1-r_{p}^{0})^{2}}{2i\epsilon_{L}^{0}} \langle\langle \Delta\epsilon_{xx} \rangle\rangle + 
	\frac{Q^{2}d \epsilon_{L}^{0} (1+r_{p}^{0})^{2} }{2iq} \langle\langle \Delta\epsilon_{zz}^{-1} \rangle\rangle}{
	1+\frac{qd (1-r_{p}^{0})}{2i\epsilon_{L}^{0}} 	\langle\langle \Delta\epsilon_{xx} \rangle\rangle 
	- \frac{Q^{2}d \epsilon_{L}^{0} (1+r_{p}^{0})}{2iq}  \langle\langle \Delta\epsilon_{zz}^{-1} \rangle\rangle},
\nonumber \\
&=& r_{p}^{0}\Bigg[1+ \nonumber \\
&&\frac{\frac{2iqd}{\epsilon_{L}^{0}-\epsilon_{S}^{0}} 
	\frac{\epsilon_{L}^{0}}{\epsilon_{S}^{0}q^{2}-\epsilon_{L}^{0}Q^{2}}
	\left(k^{2}\langle\langle \Delta\epsilon_{xx} \rangle\rangle  + Q^{2}(\epsilon_{S}^{0})^{2}
	\langle\langle \Delta\epsilon_{zz}^{-1} \rangle\rangle
	\right)}{1+\frac{qd (1-r_{p}^{0})}{2i\epsilon_{L}^{0}} 	\langle\langle \Delta\epsilon_{xx} \rangle\rangle 
	- \frac{Q^{2}d \epsilon_{L}^{0} (1+r_{p}^{0})}{2iq}  \langle\langle \Delta\epsilon_{zz}^{-1} \rangle\rangle
	}
\Bigg], \nonumber \\
\label{eq:rp-appr}
\end{eqnarray}
where $r_{p}^{0} = \frac{\epsilon_{S}^{0}q-\epsilon_{L}^{0}k}{\epsilon_{S}^{0}q+\epsilon_{L}^{0}k}$  and we have used the identity 
$(1-r_{p}^{0})^{2} = \{(1+r_{p}^{0}) (\epsilon_{L}^{0}k/\epsilon_{S}^{0}q)\}^{2}
	= r_{p}^{0}\frac{4\epsilon_{L}^{0}k^{2}}{\epsilon_{S}^{0}-\epsilon_{L}^{0}}\frac{\epsilon_{L}^{0}}{\epsilon_{S}^{0}q^{2}-\epsilon_{L}^{0}Q^{2}}$.   
From $r_{s}$ and $r_{p}$, we obtain the ellipsometry angles $\Psi$ and $\Delta$ defined as 
\begin{eqnarray}
\Psi = \tan^{-1}(\lvert r_{p}/r_{s} \rvert)
\label{eq:Psi}
\end{eqnarray}
and 
\begin{eqnarray}
\Delta = \arg(r_{p}/r_{s})~.
\end{eqnarray}
%$Q = \sqrt{\epsilon_{L}^{0}}\omega \sin\theta/c$

\section{Results: Electronic Structure \label{SCHartree}}
In this section, we present the electronic structure obtained from the solution of the  self-consistent Hartree equations, with a focus on the dependence of the subband structure and the charge distribution on the value of the polar charge $n_0$. 

A solution to the self consistent equations requires values for the static dielectric function between the different planes. We estimate these  by fitting our calculated electron density  to that obtained from DFT calculations including full lattice relaxation \cite{chen}.   We obtain $\{ \epsilon_{ii+1}\} = \{75, 350, 1000, 2000, 3000, 5000, 5000, ...\}$ where we assign 5000 for $i$ greater than 6 because our results turn out to be insensitive to the precise values of the dielectric constant, especially far from the interface, as long as these are large. The increase of the  dielectric constant with increasing distance from the interface may be understood as a nonlinear effect arising from the electric field dependence of $\epsilon$ \cite{hemberger}: near the interface the internal electric fields are large, thus reducing $\epsilon$ whereas at longer distances the fields rapidly decrease, allowing $\epsilon$ to approach the very large value appropriate to bulk STO.

The calculations of Ref.~\onlinecite{chen} were performed for a situation corresponding in our notation to polar charge equal to $0.5$ per in-plane unit cell. For smaller polar charges one expects weaker electric fields and therefore larger $\epsilon$. In all of our calculations, however, we used the same effective dielectric constant for all polar charge densities since we find that the change in the dielectric constants fitted for smaller polar charge density is not drastic and the characteristic features of electronic structure do not crucially depend on the detail of dielectric constants. Moreover, fixing $\epsilon$ allows us to isolate the effect of changing polar charge on the electronic structure. 

\begin{figure}[htbp]
\begin{center}
\includegraphics[width=1\columnwidth, angle=-0]{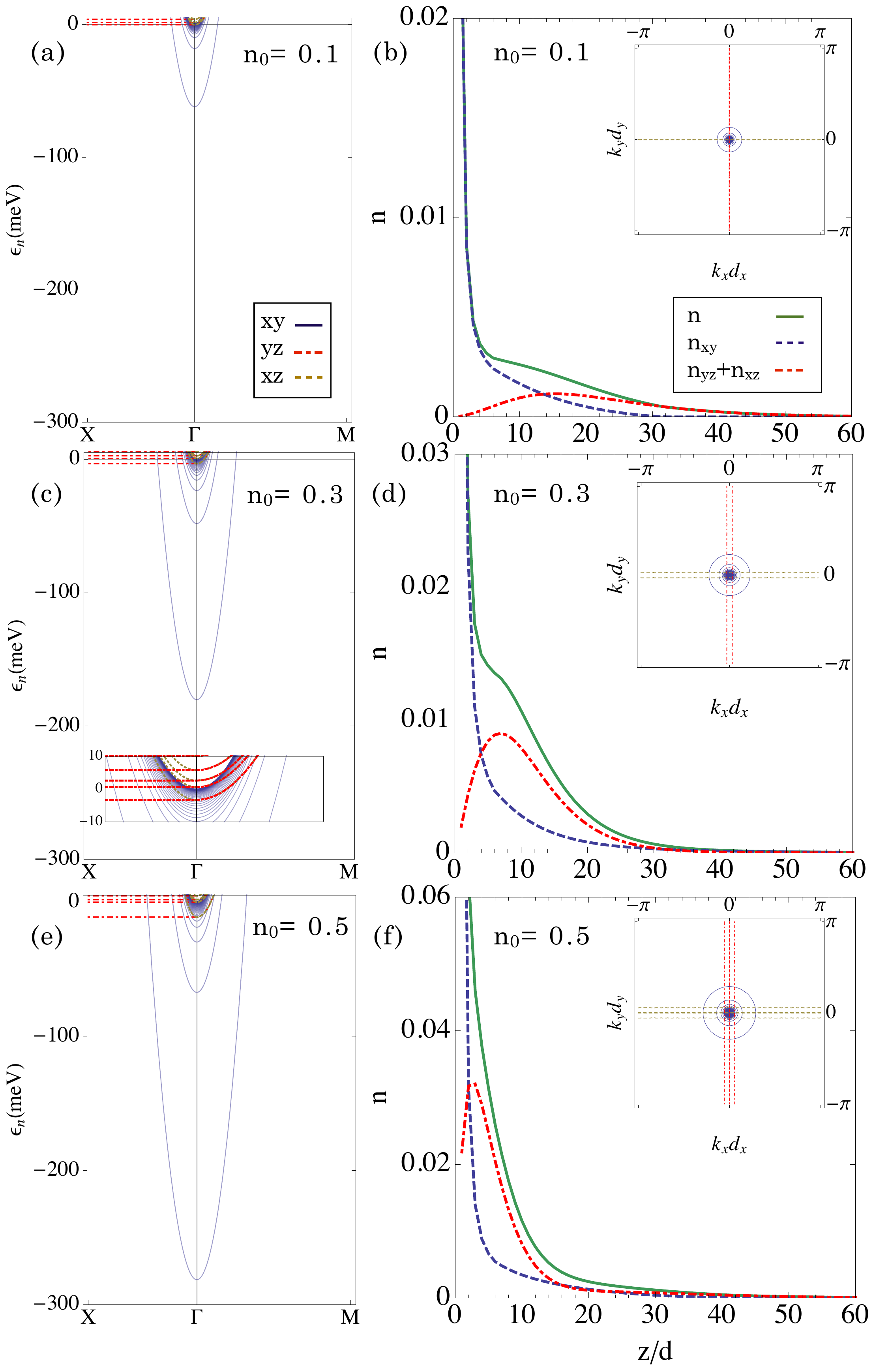}
\caption{(color online) Left panels: Electronic structure of interface electrons for the polar charge of (a) 0.1, (c) 0.3, and (e) 0.5 electrons for unit cell. Inset: magnification of lower lying $yz$ and $xz$ bands. Right panels: Charge density profiles corresponding to the three situations shown in the left panels.  Inset: Fermi surfaces for the three cases. The amount of electrons occupied in xy and $xz/yz$ bands are about 0.07 and 0.03, respectively, for the polar charge of 0.1 electrons per unit cell, 0.17 and 0.13 for 0.3 polar charge, and 0.24 and 0.26 for 0.5 polar change.}
\label{fig:2}
\end{center}
\end{figure}

Fig.~\ref{fig:2} shows the band structure and charge density distribution of interface electrons for three different polar discontinuities.  The self-consistent Hartree equation is solved for 80 TiO$_{2}$ layers and we found that as long as the number of layers was taken to be greater than 40 the results are  essentially independent of layer number. (As the number of layers goes to infinity negligibly small occupancies of higher xz/yz subband may appear.) For the $n_0=0.5$ case the results are consistent with published band structure calculations \cite{chen}. 

The right hand panels of Fig.~\ref{fig:2} show the total and orbitally resolved charge densities. As expected, the charge is confined to the near interface region and the occupations of the $xy$ and $xz/yz$ orbitals are inequivalent because of the symmetry breaking arising from the interface.  There are three spatial regimes of charge distribution: a very large peak (arising physically from the $xy$ orbitals) coming from the plane nearest the interface, an intermediate regime ($z \lesssim 30d $) with a slow spatial decay and an appreciable contribution from the $xz/yz$ orbitals (whose charge density in the very near interface regime is suppressed by wave function effects), and a far regime with a very small charge density, exponentially decaying with $z$ and indistinguishable from 0 in the Fig.~\ref{fig:2}.  The bulk of the charge is accommodated within $\sim5-10nm$ ($z\sim 10-25 d$) of the interface, in agreement with experiments. The spatial extent of the charge distribution depends on the magnitude of the polar charge, being largest for smallest polar charge and reflecting the strength of the confining potential.   For all of the cases we have examined, about $30\%$ percent of the electrons ( $50\%$ of the $xy$ electrons) are located on the first layer.

The left hand panels display the subband energies as functions of in-plane momentum in the $x$ and $y$ directions. The $d_{xz/yz}$ bands are distinguished by their lack of dispersion along $y$ and $x$ respectively. For polar charge less than $n_0\sim0.04$ (not shown) only $xy$ bands are occupied. For larger $n_0$, only one $xz$ and one $yz$ band is occupied except for a narrow range of $n_0$ very near $0.5$ where there is a very small occupancy of the second  $xz$ and $yz$ subband. Thus the $z$ dependence of the  $xz/yz$ contribution to the charge density shown in the right hand panels arises almost entirely from the $z$ dependence of the subband wave function. On the other hand a very large number of $xy$ bands are very slightly occupied, and it is these bands which give rise to the very small Thomas-Fermi tail of charge density extending very far from the interface.

\section{Ellipsometry Angles\label{Ellipsometry}}
In this section, the optical response of the interface electrons is presented in terms of ellipsometry angles, which are a convenient experimental quantity because they permit a straightforward subtraction of the substrate and overlayer effects. In particular, even in the absence of an electron gas the optical properties have features associated with reflection from the LAO-STO interface.  This structure is complicated by the very strong optical phonon of STO, which leads to  strong frequency dependence of the STO dielectric constant,   which is such that at one particular frequency the optical constants of LAO and STO nearly match and transmission through the interface becomes nearly perfect. Consideration of changes in ellipsometry angles (defined more precisely below) allows these effects to be efficiently normalized out. The characteristic features of ellipsometry angles are identified and we will relate them with physical properties of the system such as charge density, orbital occupancy, and scattering rates. 

In section II C, we derived expressions for the  reflectivity of s- and p- polarized incident light. For s-polarized light, the electric field is purely transverse and in-plane, so in the experimentally relevant long-wavelength limit,  $r_{s}$ represents the in-plane Drude response of the interface electron gas. For p-polarized incident light with both in-plane and out-of-plane components of electric field; the breaking of translation invariance in the out-of-plane direction means that charge fluctuations are induced and the longitudinal response is relevant:  the reflectivity contains both Drude and inter-band plasmonic responses. Even in the absence of interface electrons $r_s$ and $r_p$ have structure associated with the STO optical phonon. In the absence of interface electrons, forming the ratio $r_p/r_s$ removes this structure leaving a smooth behavior. Thus, by comparing the ratio $\vert r_{p}/r_{s} \vert$ with the interface electrons to the one without electrons, we can differentiate the responses of interface electrons from those from optical phonon.  Ellipsometry experiments determine $\Psi = \tan^{-1}(\vert r_{p}/r_{s} \vert)$ and $\Delta = \arg( r_{p}/r_{s} )$ simultaneously. Thus, calculating the difference $\Delta\Psi$ between the ellipsometry angles with and without interface electrons enables us to investigate the properties of that electron gas at the interface. 
 
\begin{figure}[htbp]
\begin{center}
\includegraphics[width=1\columnwidth, angle=-0]{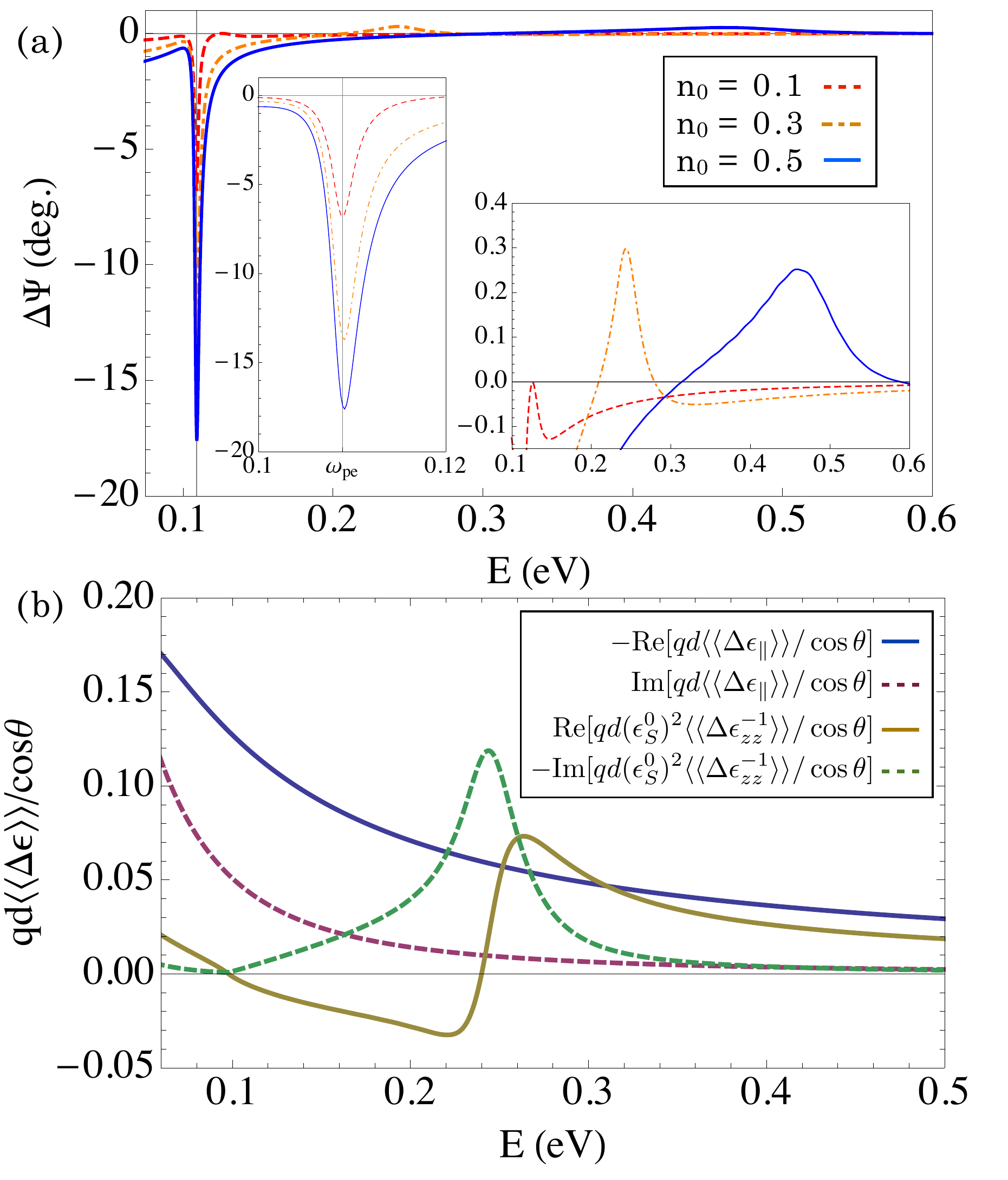}
\caption{(color online) (a) Ellipsometry angles for the polar charge of 0.1, 0.3, and 0.5 electrons per unit cell with incident angle of $75^{\circ}$. Inset (left): magnification near the plasma edge where $\text{Re}[\epsilon_{L}^{0}(\omega)]=\text{Re}[\epsilon_{S}^{0}(\omega)]$. Vertical line indicates bare interface plasma edge frequency $\omega_{pe}$. Inset (right): magnification of the higher energy peaks. (b) Real and imaginary part of the dielectric tensor contributed by interface electrons for $n_{0} = 0.3$}
\label{fig:3}
\end{center}
\end{figure}

Fig.~\ref{fig:3} presents our results as difference in  ellipsometry angles $\Delta\Psi$ with and without interface electrons   for three different polar discontinuity of 0.1, 0.3, and 0.5 electrons per unit cell. There are two main features in the ellipsometry spectra. One is a dip (highlighted in the left inset of Fig.~\ref{fig:3} (a)) approximately at the plasma edge frequency $\omega_{pe}$ where the real part of dielectric constant of STO becomes the same as the dielectric constant of LAO (assumed as the same as vacuum dielectric constant) and the other is a relatively broad peak at (highlighted in the right inset of Fig.~\ref{fig:3} (a)) high energy related to the electronic plasma excitation. The position of the high energy peak  is seen to be roughly proportional to the number of electrons in the  yz (xz) bands for the two larger charge densities. For the polar discontinuity of 0.1 the peak  position is affected by proximity to the  STO LO phonon feature. A detailed analysis of the peak position will be given later in the section. 

In order to understand the relation between ellipsometry angles and the physical properties of the system, we consider that the electronic contribution to the reflectivity is small compared with unity and thus the denominator of Eqs.~(\ref{eq:rs-appr}) and (\ref{eq:rp-appr}) are approximated as one. The approximation is valid  in broad frequency ranges satisfying both $qd \langle\langle \Delta \epsilon_{yz/xz} \rangle\rangle /\cos^{2}\theta \ll 1$ and $qd \langle\langle \Delta \epsilon_{zz}^{-1} \rangle\rangle /\cos^{2}\theta \ll 1$ since the thickness of the conducting sheet is small relative to the wavelength of light.
In this limit we have 
\begin{eqnarray}
r_{s} &\simeq&  r_{s}^{0} + \frac{iqd(1+r_{s}^{0})^{2}}{2 \epsilon_{L}^{0} \cos^{2}\theta} 
	\langle \langle \Delta \epsilon_{\parallel} \rangle \rangle~,  
\label{eq:rs-lin}\\
r_{p} &\simeq& r_{p}^{0} - \frac{iqd (1-r_{p}^{0})^{2}}{2 \epsilon_{L}^{0}} \bigg[
	\langle \langle \Delta \epsilon_{\parallel} \rangle \rangle \nonumber \\
&&+ \frac{\epsilon_{L}^{0}\sin^{2}\theta}{\epsilon_{S}^{0}-\epsilon_{L}^{0}\sin^{2}\theta} 
	(\epsilon_{S}^{0})^{2} \langle \langle \Delta \epsilon_{zz}^{-1} \rangle \rangle~,
	\bigg]~.
\label{eq:rp-lin}
\end{eqnarray}
where we define $\langle \langle \Delta \epsilon_{\parallel} \rangle \rangle \equiv \langle \langle \Delta \epsilon_{xx} \rangle \rangle = \langle \langle \Delta \epsilon_{yy} \rangle \rangle$ using the cubic symmetry. 

With the approximated expressions for $r_{s}$ and $r_{p}$ we consider two simplifying limits. First, we consider that the electronic contributions to the reflection relative to that from the mismatch of the two bulk dielectric constants is small and only take into account the effect of interface electrons up to the linear order. Then the ratio $r_{p}/r_{s}$ becomes 
\begin{eqnarray}
\frac{r_{p}}{r_{s}} &=& \frac{r_{p}^{0}}{r_{s}^{0}} - \frac{iqd}{2 \epsilon_{L}^{0}} \Big[
\left(\frac{(1-r_{p}^{0})^{2}}{r_{s}^{0}} +\frac{r_{p}^{0}(1+r_{s}^{0})^{2}}{(r_{s}^{0})^{2}\cos^{2}\theta} \right)
	 \langle \langle\Delta \epsilon_{\parallel} \rangle \rangle \nonumber \\
&& + \frac{(1-r_{p}^{0})^{2}}{r_{s}^{0}}  \frac{\epsilon_{L}^{0}\sin^{2}\theta}{\epsilon_{S}^{0}-\epsilon_{L}^{0}\sin^{2}\theta} 
	(\epsilon_{S}^{0})^{2} \langle \langle \Delta \epsilon_{zz}^{-1} \rangle \rangle 
\Big]~.
\label{eq:appr-rp-ov-rs}
\end{eqnarray}
We expect that at high frequencies ($\omega \gg 1/\tau$) $\langle \langle \Delta\epsilon_{\parallel} \rangle\rangle$, essentially, Drude conductivity becomes small whereas $(\epsilon_{S}^{0})^{2}\langle \langle \Delta \epsilon_{zz}^{-1} \rangle\rangle$ has a peak at an appropriate plasma frequency. In the low frequency limit ($\omega \ll 1/\tau$), the contribution from $\langle \langle \Delta\epsilon_{\parallel} \rangle\rangle$ dominates due to the $1/\omega$ factor in $\langle \langle \Delta\epsilon_{\parallel} \rangle\rangle$ whereas  $(\epsilon_{S}^{0})^{2}\langle \langle \Delta \epsilon_{zz}^{-1} \rangle\rangle$ is finite as $\omega \rightarrow 0$. 

In the opposite limit, which may occur at the $\omega_{pe}$ at which $\text{Re}[\epsilon_{L}^{0} ]= \text{Re}[\epsilon_{S}^{0}]$, we have vanishing $r_{s}^{0}$ and $r_{p}^{0}$ if $\text{Im} [\epsilon_{S}^{0}]$ is small. If the contribution from the interface electrons satisfies $\text{Im}[\epsilon_{S}^{0}] \ll qd\langle \langle \Delta\epsilon_{\parallel} \rangle\rangle/\cos^{2}\theta$, $qd( \epsilon_{S}^{0})^{2}\langle \langle \Delta \epsilon_{zz}^{-1} \rangle\rangle/\cos^{2}\theta$, up to linear order, $r_{p}/r_{s}$ can be written as
\begin{eqnarray}
\frac{r_{p}}{r_{s}} &=& -\cos^{2}\theta - 
\sin^{2}\theta (\epsilon_{S}^{0})^{2} \langle \langle \Delta \epsilon_{zz}^{-1} \rangle \rangle /\langle \langle \Delta\epsilon_{\parallel} \rangle\rangle 
\nonumber \\
&-& \frac{2i\epsilon_{L}^{0}\cos^{2}\theta}{qd  \langle \langle \Delta\epsilon_{\parallel} \rangle\rangle} [
r_{p}^{0}+r_{s}^{0}\cos^{2}\theta \nonumber \\
&&+  r_{s}^{0}\cos^{2}\theta(\epsilon_{S}^{0})^{2} \langle \langle \Delta \epsilon_{zz}^{-1} \rangle \rangle /\langle \langle \Delta\epsilon_{\parallel} \rangle\rangle 
]~,
\label{eq:rp-ov-rs-appr2}
\end{eqnarray} 
where the contribution from the out-of-plane excitation is expressed as a ratio of out of plane to in-plane dielectric tensor.

In order to analyze the feature in the ellipsometry angles, we first examine the dielectric tensor $\langle\langle \Delta \epsilon_{\parallel}(\omega_{pe}) \rangle\rangle$ and $(\epsilon_{S}^{0})^{2} \langle\langle \Delta\epsilon_{zz}^{-1}(\omega_{pe}) \rangle\rangle$ multiplied by $qd/\cos\theta$. Fig.~\ref{fig:3} (b) shows the real and imaginary part of $\langle\langle \Delta \epsilon_{\parallel} \rangle\rangle$ and $( \epsilon_{S}^{0})^{2} \langle\langle \Delta\epsilon_{zz}^{-1} \rangle\rangle$ for $n_{0}=0.3$. First of all, we can see that overall magnitude of dielectric tensor is well below 1 so the approximation used in Eqs.~(\ref{eq:rs-lin}) and (\ref{eq:rp-lin}) is valid except very near grazing incidence $(\theta \sim 90^{\circ})$. Moreover, we find that the in-plane dielectric tensor dominates especially near the STO plasma edge $\omega_{pe}\sim 0.11\; eV$, which is also common for $n_{0}=0.1$ and $n_{0}=0.5$ cases (not shown). Near $\omega_{pe}$, both the real and imaginary part of $\epsilon_{S}^{0}$ is small ($\text{Re}[\epsilon_{S}^{0}]\simeq \epsilon_{L}^{0}\sim 1, \text{Im}[\epsilon_{S}^{0}]\sim 0.1$) and in this case the density correlation function $\hat \Pi$ is roughly proportional to the inverse of the Coulomb matrix. Then the out-of-plane dielectric tensor can be written as $-qd (\epsilon_{S}^{0})^{2} \langle\langle \Delta \epsilon_{zz}^{-1} \rangle\rangle \sim qL$ which is much smaller than one. Thus, we expect that the dip at $\omega_{pe}$ is due mostly to the in-plane conductivity. Since the real part of the bare dielectric constant of STO and LAO coincides at $\omega_{pe}$ and with its small imaginary part we can expand Eqs.~(\ref{eq:rs-lin}) and (\ref{eq:rp-lin}) up to the linear order of $\text{Im}[\epsilon_{S}^{0}]$ and electronic dielectric tensor 
\begin{eqnarray}
r_{s}\vert_{\omega=\omega_{pe}} &=&\left[
  -\frac{i\text{Im}[\epsilon_{S}^{0}]}{4 \epsilon_{L}^{0} \cos^{2}\theta} + \frac{iqd\langle \langle \Delta\epsilon_{\parallel} \rangle\rangle}{2\epsilon_{L}^{0} \cos^{2}\theta} 
  \right]_{\omega=\omega_{pe}} \\ 
r_{p}\vert_{\omega=\omega_{pe}}  &=&  \left[ \frac{i(2\cos^{2}\theta - 1)\text{Im}[\epsilon_{S}^{0}]}{4 \epsilon_{L}^{0}cos^{2}\theta}- \frac{iqd\langle \langle \Delta\epsilon_{\parallel} \rangle\rangle}{2\epsilon_{L}^{0}}
\right]_{\omega=\omega_{pe}}~, \nonumber \\
\end{eqnarray}
where we ignore the out-of-plane dielectric tensor and have used the bare reflectivity at $\omega_{pe}$
\begin{eqnarray}
r_{s}^{0} \vert_{\omega=\omega_{pe}} &\simeq& -\frac{i\text{Im}[\epsilon_{S}^{0}]}{4 \epsilon_{L}^{0} \cos^{2}\theta}~, \\
r_{p}^{0}\vert_{\omega=\omega_{pe}} &\simeq& \frac{i(2\cos^{2}\theta - 1)\text{Im}[\epsilon_{S}^{0}]}{4 \epsilon_{L}^{0}cos^{2}\theta}~.
\end{eqnarray}
Since $\text{Im}[\epsilon_{S}^{0}]$ is in the same order of $qd \langle \langle \Delta\epsilon_{\parallel} \rangle\rangle$,  Eqs.~(\ref{eq:appr-rp-ov-rs}) and (\ref{eq:rp-ov-rs-appr2}) are not applicable. However, we can get some insight by evaluating the magnitude of $r_{p}/r_{s}$
\begin{widetext}
\begin{eqnarray}
\bigg\vert\frac{r_{p}}{r_{s}} \bigg\vert_{\omega=\omega_{pe}} = \left[\frac{\left((2\cos^{2}\theta-1)\text{Im}[\epsilon_{S}^{0}] 
	+ \frac{2q_{pe}d\Omega_{\parallel}^{2} \cos^{2}\theta}{\omega_{pe}^{2}+1/\tau^{2}} \right)^{2}+ 
	\left(\frac{2q_{pe}d \Omega_{\parallel}^{2}\cos^{2}\theta/\tau}{\omega_{pe}^{3}+ \omega_{pe}/\tau^{2}} \right)^{2}}{
\left(\text{Im}[\epsilon_{S}^{0}] 
	+ \frac{2q_{pe}d\Omega_{\parallel}^{2} }{\omega_{pe}^{2}+1/\tau^{2}} \right)^{2}+ \left(\frac{2q_{pe}d \Omega_{\parallel}^{2}/\tau}{\omega_{pe}^{3}+ \omega_{pe}/\tau^{2}} \right)^{2}} \right]^{1/2}~,
\label{eq:mag-lin-rp-ov-rs}
\end{eqnarray}
\end{widetext}
where $q_{pe}$ is defined as the wavevector at $\omega_{pe}$.

\begin{figure}[htbp]
\begin{center}
\includegraphics[width=0.9\columnwidth, angle=-0]{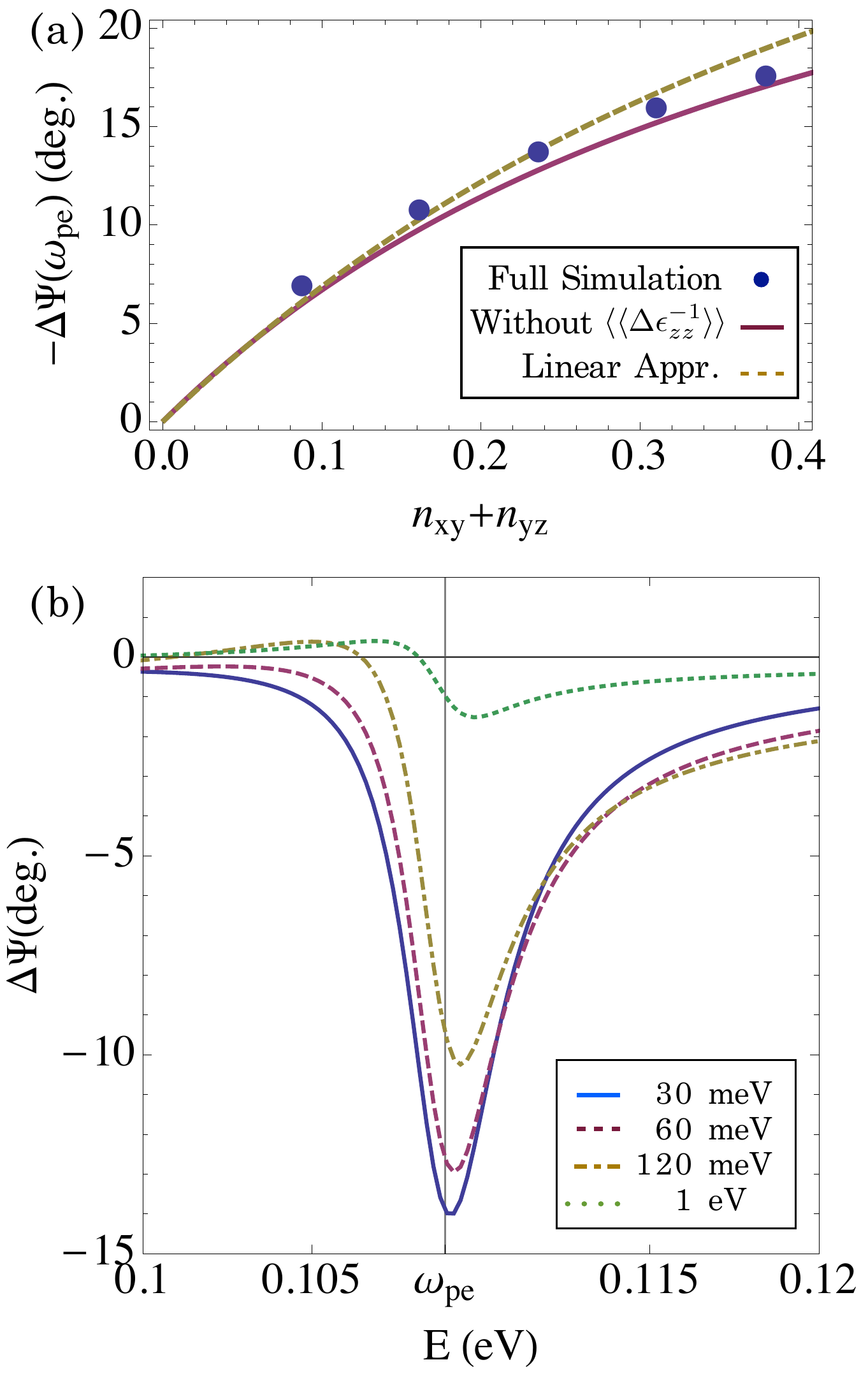}
\caption{(color online) The structure of ellipsometry angles near $\omega_{pe}$.  (a) Magnitude of the dip at the plasma edge frequency as a function of $n_{xy}+n_{yz}$. The dots are from the simulated ellipsometry spectra and the solid line is obtained from the Eqs.~(\ref{eq:rs-appr}) and (\ref{eq:rp-appr}) while ignoring the $\langle\langle \Delta \epsilon_{zz}^{-1}\rangle\rangle$. The dashed line is obtained by Eq.~(\ref{eq:mag-lin-rp-ov-rs}). (b) Comparison of ellipsometry  angles with the different scattering rates for polar charge of 0.3 electrons per unit cell around the plasma edge.}
\label{fig:4}
\end{center}
\end{figure}

For $\theta \leq 45^{\circ}$, we can show that $d\vert r_{p}/r_{s}\vert/ d\Omega_{\parallel}^{2} > 0$ regardless of the value of $\Omega_{\parallel}^{2}$ and thus we expect that  $\vert r_{p}/r_{s} \vert - \vert r_{p}^{0}/r_{s}^{0}  \vert$ at $\omega_{pe}$ is always positive and monotonically increase as the density $n_{xy}+n_{yz}$ increases. On the other hand, for $\theta > 45^{\circ}$,  $d\vert r_{p}/r_{s}\vert/ d\Omega_{\parallel}^{2}$  becomes negative for small $\Omega_{\parallel}^{2}$ and change its sign around 
\begin{eqnarray}
\Omega_{\parallel c}^{2}\simeq\frac{(1-2\cos^{2}\theta)(\omega_{pe}^{2}+1/\tau^{2})\text{Im}[\epsilon_{S}^{0}]}{2q_{pe}d \cos^{2}\theta}
\end{eqnarray}
assuming $\omega_{pe} \gg 1/\tau$. Thus, as we increase the polar discontinuity, $\Delta\Psi$ will increase monotonically for $\theta \leq 45^{\circ}$ and for $\theta > 45^{\circ}$, it will decrease until $\Omega_{\parallel}^{2}\simeq\Omega_{\parallel c}^{2}$. 
We can also evaluate the effect of varying the scattering rate $1/\tau$. In the regime $1/\tau \ll \omega_{pe}$, increasing the scattering rate is equivalent to changing $\Omega_{\parallel}^{2} \rightarrow  \Omega_{\parallel}^{2} (1-1/(\omega_{pe}\tau)^{2})$. Therefore, the magnitude of $\Delta\Psi$ will decrease as the scattering rate increases as long as $1/\tau \lesssim \omega_{pe}$. Moreover, Eq.~(\ref{eq:mag-lin-rp-ov-rs}) shows that with a given incident angle and bare dielectric constants, we can deduce the information that $\Omega_{\parallel}^{2}$ is proportional to $n_{xy}+n_{yz}$ and the scattering rate $1/\tau$. Fig.~\ref{fig:4} (a) shows the ellipsometry angles comparing our approximated expression (Eq.~(\ref{eq:mag-lin-rp-ov-rs})) and full simulation results based on Eqs.~(\ref{eq:rs-appr}) and (\ref{eq:rp-appr}). We can see that the linear approximation is valid even for high occupation of $xy$ and $yz$ bands and also find that ignoring the contribution from $\langle \langle \Delta\epsilon_{zz}^{-1} \rangle\rangle$ gives results closed to the full simulation values. In Fig.~\ref{fig:4} (b) the ellipsometry angles near $\omega_{pe}$ is investigated. As expected the reduction of the dip value is observed and additionally for high scattering rate ($1/\tau \sim \omega_{pe}$) we find that a small peak emerges in the lower energy side of $\omega_{pe}$.

We also investigate the peak visible in the higher frequency range on Fig.~\ref{fig:3} (a). In this case  we can apply Eq.~(\ref{eq:appr-rp-ov-rs}),  since  both $r_{s}^{0}$ and $r_{s}^{0}$ give dominant contribution to the reflectivity for the frequency away from $\omega_{pe}$. In this region, the bare reflectivity is mostly real and both it and the in-plane dielectric tensor (as seen in Fig.~\ref{fig:3}(b)) have smooth behavior. Thus we can expect that the peak in the ellipsometry angles are from the out-of-plane dielectric tensor $\langle \langle \Delta\epsilon_{zz}^{-1} \rangle\rangle$. To see this, we rewrite Eq.~(\ref{eq:appr-rp-ov-rs}) up to the linear order of electronic dielectric tensor as 
\begin{eqnarray}
&&\left\vert \frac{r_{p}}{r_{s}} \right\vert - \left\vert \frac{r_{p}^{0}}{r_{s}^{0}} \right\vert  \nonumber \\
&&\simeq \frac{qd}{2\epsilon_{L}^{0}} \left\vert \frac{r_{p}^{0}}{r_{s}^{0}} \right\vert \bigg[
\left( \frac{(1-r_{p}^{0})^{2}}{r_{p}^{0}} + \frac{(1+r_{s}^{0})^{2}}{r_{s}^{0}} \right) \text{Im}[\langle\langle \Delta \epsilon_{\parallel} \rangle\rangle] \nonumber \\
&&+\frac{(1-r_{p}^{0})^{2}}{r_{p}^{0}}  \frac{\epsilon_{L}^{0}\sin^{2}\theta}{\epsilon_{S}^{0}-\epsilon_{L}^{0}\sin^{2}\theta} 
\text{Im}[(\epsilon_{S}^{0})^{2} \langle \langle \Delta \epsilon_{zz}^{-1} \rangle \rangle] 
\bigg]~,
\label{eq:appr-mag-rp-ov-rs}
\end{eqnarray}
where we ignore the imaginary part of $\epsilon_{S}^{0}$. Given that reflectivity are not sensitive to the frequency and both the sign of $r_{s}^{0}$ and $r_{p}^{0}$ are negative, we can see that the effect of in-plane dielectric tensor is to give smooth background $\propto -\Omega_{\parallel}^{2} /(\omega^{2}+ 1/\tau^{2})$ and the peak is mostly due to the imaginary part of the out-of-the-plane dielectric tensor. The relation can also seen in Fig.~\ref{fig:3} (b) where the peak energy of $-\text{Im}[qd\langle\langle \Delta \epsilon_{zz}^{-1} \rangle \rangle]$ coincides with the peak in the right inset in Fig.~\ref{fig:3} (a) for $n_{0} = 0.3$. 

In Sec. \ref{Model} (b), we express the out-of-plane dielectric tensor with the density correlation function of $xz/yz$ electrons (Eq.~(\ref{eq:Pi-appr})). Since the imaginary part of the density correlation function represents the plasmon excitations, we further investigate the density correlation function $\hat\Pi$ (Eq.~(\ref{eq:PI-mm})). We can write the inverse of the $\hat\Pi$ as
\begin{widetext}
\begin{eqnarray}
\Pi_{mm^{\prime}}^{-1}(\omega) &=& \frac{\omega(\omega+i/\tau)\delta_{mm^{\prime}} 
- \Delta_{mm_{0}}^{2}\delta_{mm^{\prime} }-4n_{m_{0}}\sqrt{\Delta_{mm_{0}}}v_{mm^{\prime}}\sqrt{\Delta_{mm_{0}}}
}{ 4n_{m_{0}}\sqrt{\Delta_{mm_{0}}}\sqrt{\Delta_{m^{\prime}m_{0}}} }
\nonumber \\
&=& \frac{\sum_{m_{1}}U^{\ast}_{mm_{1}}\left[
\omega(\omega+i/ \tau) - g_{m_{1}}
\right] U_{m_{1}m^{\prime}}}{4n_{m_{0}} \sqrt{\Delta_{mm_{0}}}\sqrt{\Delta_{m^{\prime}m_{0}}}} ~,
\end{eqnarray}
\end{widetext}
where $U_{mm^{\prime}}$ is a frequency independent unitary matrix satisfying 
$U^{\ast}_{mm_{1}} g_{m_{1}} U_{m_{1}m^{\prime}} =   \Delta_{mm_{0}}^{2}\delta_{mm^{\prime} } + 4n_{m_{0}}\sqrt{\Delta_{mm_{0}}} v_{mm^{\prime}}\sqrt{\Delta_{m^{\prime}m_{0}}}$ 
using that the bare dielectric constant of STO does not change significantly at frequencies above the STO phonon frequency and is treated as a constant. 
Thus, we can get a simplified expression for  $\langle \langle \Delta \epsilon_{zz}^{-1} \rangle \rangle$ as 
\begin{eqnarray}
(\epsilon_{S}^{0})^{2} \langle \langle \Delta\epsilon^{-1}_{zz}(\omega) \rangle \rangle
&=& \sum_{m} \frac{\Omega_{\perp}^{2} \vert f_{m} \vert^{2}}{\omega(\omega + i/ \tau) - g_{m}}~,
\label{eq:diag-Pi}
\end{eqnarray}
where $f_{m} = \sum_{m^{\prime}}
	U_{m m^{\prime} }\sqrt{2\Delta_{m^{\prime}m_{0}}/t} \langle m^{\prime} \vert z \vert m_{0}\rangle $.

\begin{figure}[htbp]
\begin{center}
\includegraphics[width=1\columnwidth, angle=-0]{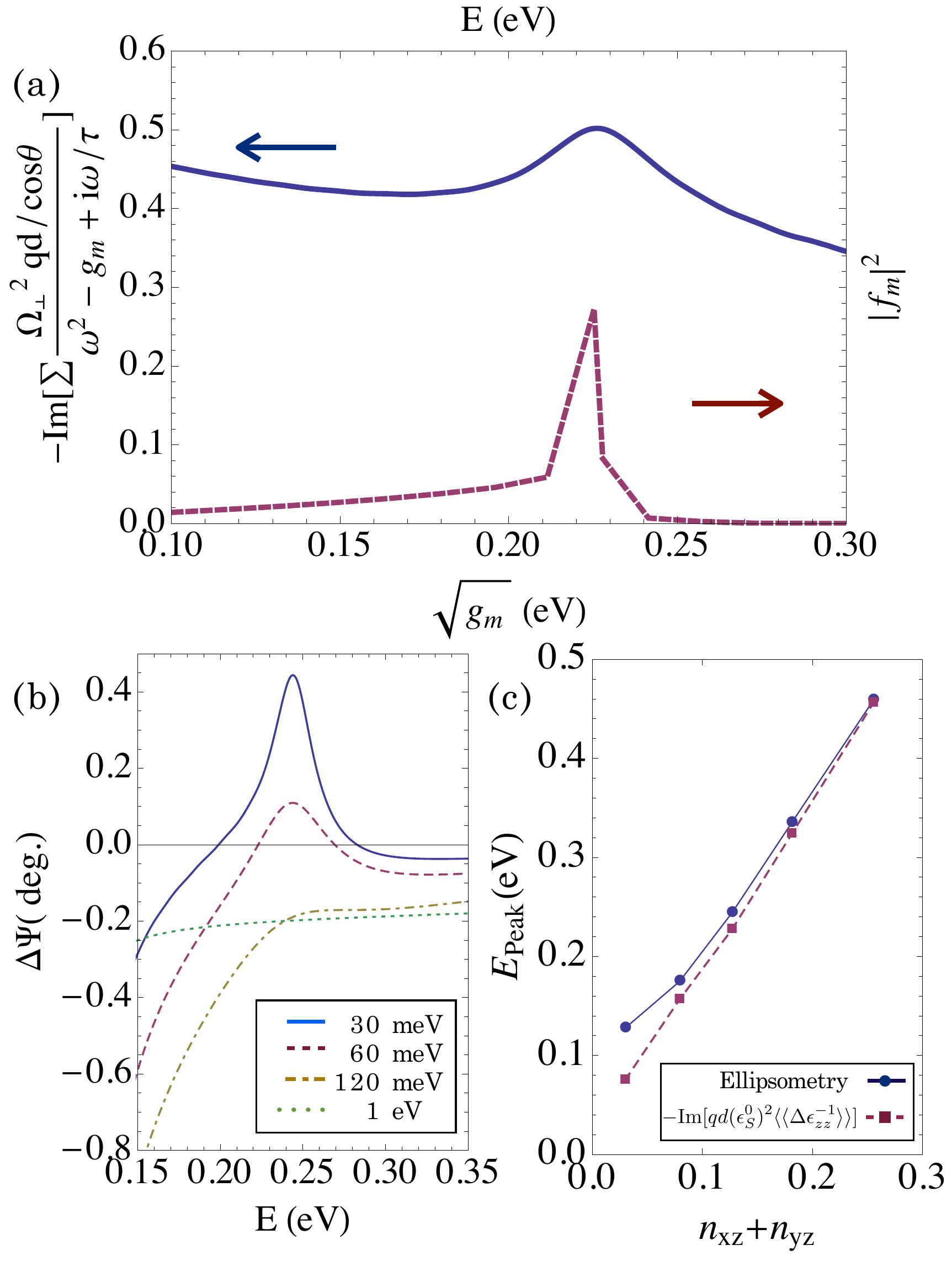}
\caption{(color online) (a) Diagonalized matrix elements (red dashed line) and density correlation function (blue line) for $n_{0} = 0.3$. 
(b) Comparison of ellipsometry  angles with the different scattering rates for polar charge of 0.3 electrons per unit cell around plasmon peak.
(c) Peak positions in ellipsometry spectra compared with the peak in the out-of-plane dielectric tensor. }
\label{fig:5}
\end{center}
\end{figure}

With Eq.~(\ref{eq:diag-Pi}) we interpret the structure of the out-of-plane dielectric constant obtained by RPA. Without screening we have poles in $\hat \Pi$ at energies corresponds to interband transition between the singly occupied state of $xz$ ($yz$) band and the unoccupied states. The matrix elements are monotonically decreasing as increasing inter-band energy. However including the density response (screening) of the electron, expressed as Coulomb matrix $v_{mm^{\prime}}$, shifts the poles and mixes the matrix elements where shifted pole frequencies represent plasmon excitations related to the longitudinal oscillation of the $xz/yz$ electrons and their oscillator strength are determined by the matrix elements $f_{m}$ which are linear combinations of dipole matrix elements.  Thus, we can interpret the peak of the ellipsometry angle as  arising from plasmon excitations that are determined by two factors: how the eigenvalues $g_{m}$ are close to each other and the oscillator strength $f_{m}$ which are determined by band structure of the $yz/xz$ electrons. In Fig.~\ref{fig:6} (a) shows the diagonalized density correlation functions and its oscillator strength. We can see that the dominant contribution is from the oscillator strength while the eigenvalue effects are relatively small. The rapid decrease of $\vert f_{m}\vert^{2}$ for $\sqrt{g_{m}} \gtrsim 0.25 eV$ gives the asymmetric peak shape. The relation between plasma poles and the scattering rate is also investigated in Fig.~\ref{fig:6} (b). For $1/\tau \ll E_{Peak}$ defined as the energy of plasma peak (Fig.~\ref{fig:6} (c)) the plasma poles are well defined and the scattering rate can be seen as a broadening in the plasma poles and it is expected that the peak height scales as $\Omega_{\perp}^{2}\tau/\omega$. For $1/\tau \gg E_{Peak}$, the plasma poles no longer exist and the plasma peak vanishes. The plasma peak estimated by Eq.~(\ref{eq:diag-Pi}) is compared with ellipsometry data in Fig.~\ref{fig:6} (c) where for high occupation, they agrees with each other and the difference is increases for smaller occupation since in Eq.~(\ref{eq:diag-Pi}), we use dielectric constant of STO in the limit of $\omega \rightarrow \infty$. As the plasmon peak energy decreases, the bare dielectric constant of STO deviates significantly from its limiting value and near $\omega_{pe}$ the approximation used in Eq.~(\ref{eq:appr-rp-ov-rs}) is not valid any more.

We compare the position of the plasmon peaks with a simple estimation of the free electron plasma frequency $\Omega_{p}^{0} = \sqrt{\frac{8\pi e^{2 }t (n_{xz}+n_{yz}) }{\epsilon^{0}_{S}(\infty)L^{\ast}} }$ associated with an electron gas of volume density corresponding to our total areal density of interface xz/yz electrons spread over a length $L^\ast$ which we take to be twice the peak position of the $xz/yz$ density profile. The estimated plasma frequency for the polar charge density of $\{0.1, 0.3, 0.5\}$ are $\{79, 237, 512\}$ $meV$ which are fairly close to the calculated  plasmon peak position. The reasonable agreement between calculation and free electron model is expected  since the binding energy of the $xz/yz$ electrons is about two order of magnitude  smaller than the plasma frequency.

In Sec. \ref{Model} (c) presents detailed formulas for the reflectivity by taking the long-wavelength approximation to the non-local dielectric tensor. The final results can be understood as implying that the interface system is described by an anisotropic dielectric tensor with components 
\begin{eqnarray}
\bar\epsilon_{\parallel}(\omega) &\simeq& \epsilon_{S}^{0} - 2ikd\langle \langle \Delta \epsilon_{\parallel}(\omega)  \rangle \rangle \nonumber \\
\bar \epsilon_{zz}(\omega) &\simeq& \epsilon_{S}^{0} + 2 i kd  (\epsilon_{S}^{0})^{2} \langle \langle \Delta \epsilon_{zz}^{-1}(\omega) \rangle \rangle~.
\end{eqnarray}
The complete expression of the effective dielectric tensor and its characteristic features are presented in Appendix B. We note that the dielectric tensor depends on the incident angle through $kd$. The effective dielectric tensor allows us to have an alternative description of the interface electrons system.

In this section, we investigated the features in the ellipsometry measurements. By taking simplifying limits we are able to find the physical origin of the features and relate them with the physical quantities of interface electrons. Moreover we find the effective expression in terms of local dielectric tensor that may be helpful to provide another way to understand the behavior of interface electrons. Although the mapping between the plasmon peak and occupation of $xz/yz$ elections and its scattering rates depends on the detail of band structure,  combined with the information of the dip feature our results may allow us to reasonably estimate the physical properties of the interface electrons.

\section{Summary\label{Conclusion}}
We have presented an  analysis of the subband structure and optical response of an electron gas confined at a polar interface. We assumed that host material is similar to SrTiO$_3$ in having a conduction band derived from $t_{2g}$ symmetry d-orbitals and possessing a very anisotropic dispersion and that the interface could be described as a charged sheet with areal charge density $n_0$ which is a parameter of the model. We used a self-consistent Hartree approximation to treat the Coulomb confinement and computed the dielectric tensor within the RPA approximation.  Our model involves several approximations. The most crucial of these is the neglect of mixing among the three $t_{2g}$-derived bands.  Actual band-mixing effects arise from spin orbit coupling and further-neighbor hopping and are  small in the situation we consider\cite{Zhong12}. In particular, as discussed at length above, the response to in-plane electric fields is determined mainly by the number of carriers free to move parallel to the plane of the interface while the plasmon frequencies have to do with the total density of carriers with significant out-of-plane dispersion; neither of these is affected much by the small interband couplings we have neglected. The neglect of band-mixing crucially simplifies the calculations, while the flexibility of our model-system formalism means that a wide range of situations can be studied and the sensitivity of the results to effects such as the total polar charge (controlled in part by defects which may be hard to model ab-initio) leads to increased insight.  

One crucial finding (implicit in previous work but apparently not remarked upon explicitly in the literature) is that the majority  of the induced charge resides in $xy$-derived bands confined very closely (within 2 unit cells) of the interface. The charge density at farther distances has two components, with somewhat different distance dependences. One component arises from the lowest $xz/yz$ subband, which is in most circumstances the only $xz/yz$ subband to be occupied. The spatial decay of the charge density in this subband is determined by the subband wave function. The second component arises from a ``Thomas Fermi tail'' of electrons in $xy$-derived bands, with occupancy controlled by the self-consistent screening of the interface potential. The charge density in this Thomas-Fermi tail is very small, but because of the very large dielectric constant of SrTiO$_3$ extends a very long distance from the interface. 

By use of our calculated subband structure, the RPA approximation and a Greens function formalism for the light propagation in an inhomogeneous situation we calculated the optical response of the interface system. Following the experimental literature \cite{dubroka} we presented the results in term of changes in ellipsometry angles between the system with and without the interface electron gas. We find that the frequency dependent change in the ellipsometry angle involves two important structures: a dip, at a frequency corresponding to the matching dielectric constants of SrTiO$_3$ and LaAlO$_{3}$ and a peak at a higher frequency. The frequency at which the dip occurs is controlled by details of the bulk STO and LAO optical responses and is not interesting. The magnitude of the dip was shown to be controlled by the in-plane conductivity. The peak in higher energy is in essence a c-axis  plasmon excitation of the interface electrons.  We presented simple estimates allowing extraction of these parameters from the data without performing an elaborate fit of our calculations. 

Our results show how ellipsometry experiments can be a generally useful tool in the analysis of the electronic properties of oxide interfaces. Given the angle of the incident light and the bare dielectric constants of STO and LAO, our results shows that the location of the plasmon peak and its height are easily mapped to the density $n_{yz}$ and the relaxation time $\tau$ through Eqs.~(\ref{eq:Psi}) and (\ref{eq:appr-mag-rp-ov-rs}). Moreover with the magnitude of dip at $\omega_{pe}$, and obtained $\tau$  we can also obtain $n_{xy}+n_{yz}$ (Eq.~( \ref{eq:mag-lin-rp-ov-rs})). Thus the information about the orbital occupancy and disorders can be inferred. It has been reported the discrepancy between the number of carriers inferred from transport measurments \cite{thiel} and the number expected from the polar catastrophe scenario \cite{Mannhart08,Bristowe09,Schlom10} or measured in high energy photoemission experiments \cite{Sing09} and the mechanism that cause the discrepancy is not clear. Our results may provide insight to this by the comparison with the recent ellipsometry measurements\cite{dubroka} that shows consistent dip and peak features.

\begin{acknowledgements}
This work is supported by  DOE ER-046169. We thank A. Dubroka, C. Bernhard,  S. Thiel, and J. Mannhart for very helpful discussions and providing unpublished data.
\end{acknowledgements}

\appendix
\section{Electromagnetic Green Function}
We introduce the electromagnetic Green function that is a solution of  Eq.~(\ref{eq:GFeq}) following the derivation by Bagchi {\it et al.}\cite{Bagchi79}. For $s-$ and $p-$ polarization, the explicit form of the Green function satisfying proper boundary conditions is presented. 

For a s-polarized light, the electric field is perpendicular to the plane of incident which is assumed to be in the $xz$ plane. Thus, the electric field is only in the $y$ direction. Assuming $\mathbf{E} \propto \hat y e^{iQx}E_{y}(z)$ we can write
\begin{eqnarray}
-\nabla \times (\nabla\times \mathbf{E} ) &=& \nabla^{2}\mathbf{E} - \nabla(\nabla \cdot \mathbf{E})  \nonumber \\
&=& \hat y (-Q^{2} +  \frac{ \partial^{2} }{\partial z^{2}})E_{y}(z)e^{iQx}~.
\end{eqnarray}
Then Eq.~(\ref{eq:deq}) becomes 
\begin{eqnarray}
&&\left[ \frac{ \partial^{2} }{\partial z^{2}} - Q^{2} \right] E_{y}(z) + \frac{\omega^{2}}{c^{2}} \epsilon^{0}_{yy}(z)E_{y}(z) \nonumber \\
&&= -\frac{\omega^{2}}{c^{2}} \int dz_{1} \Delta\hat \epsilon(z,z_{1}) E_{y}(z_{1})~
\end{eqnarray}
and the equation for the Green function is 
\begin{equation}
\left[ \frac{ \partial^{2} }{\partial z^{2}} -Q^{2} \right] G_{yy}(z,z_{1}) 
	+  \frac{\omega^{2}}{c^{2}} \epsilon^{0}(z)G_{yy}(z,z_{1}) = \delta(z-z_{1})~.
\label{eq:diffGyy}
\end{equation}
We now claim that the Green function is following form: 
\begin{eqnarray}
G_{yy}(z,z^{\prime}) &=& \beta [\Theta(z-z^{\prime}) U_{y}(z)V_{y}(z^{\prime}) \nonumber \\ 
&&+ \Theta(z^{\prime}-z)V_{y}(z)U_{y}(z^{\prime})]
\label{eq:Gyy}
\end{eqnarray}
where
\begin{eqnarray}
&&U_{y}(z) = \bigg\{
\begin{array}{cc}
e^{iqz}+ r_{s}^{0}e^{-iqz} & z < 0 \\
(1+r_{s}^{0})e^{ikz}& z > 0
\end{array}~, 
\end{eqnarray}
\begin{eqnarray}
&&V_{y}(z)= \bigg\{
\begin{array}{cc}
(1-r_{s}^{0})e^{-iqz}  & z < 0 \\
e^{-ikz}- r_{s}^{0}e^{ikz}& z > 0
\end{array}~,   \label{eq:UyVy}
\end{eqnarray}
\begin{eqnarray}
&&\beta = \frac{1}{2i k(1+r_{s}^{0})} = \frac{1}{2iq(1-r_{s}^{0})} = \frac{q+k}{4i qk}~, \\
&&r_{s}^{0} = \frac{q-k}{q+k} ~,\\
&& q = \frac{\sqrt{\epsilon^{0}_{L}}\omega\cos\theta}{c },~ k=\frac{\omega\sqrt{\epsilon_{S}^{0}-\epsilon_{L}^{0}\sin\theta}}{c}~,
\end{eqnarray}
and show that it satisfies Eq.~(\ref{eq:diffGyy}) and proper boundary conditions. 

We can see that each of $U_{y}$ and $V_{y}$ satisfies the homogeneous wave equation (Eq.~(\ref{eq:E0})). Therefore from Eq.~(\ref{eq:Escf}), we expect that in the limit of $z \rightarrow -\infty$, the asymptotic electric field becomes  $E_{y}(z)\propto e^{-ikz}$, which is consistent with the expression of the self-consistent equation in Eq.~(\ref{eq:scf-spol}). Moreover, the continuity in $U_{y}$ and $V_{y}$ across the $z=0$ ensures that the $E_{y}$ is continuous across the interface. The interpretation of the Green function follows. For example, in the case of $z,z^{\prime}>0$ the Green function is 
\begin{equation}
G_{yy}(z,z^{\prime}) = \frac{1}{2ik}
	\left( e^{ik\vert z-z^{\prime}\vert} - r_{s}^{0}e^{ik(z+z^{\prime})} \right) ~.
\end{equation}
The first term corresponds to the propagation of the wave from $z^{\prime}$ to $z$ and the second term is the wave reflected at the interface
and propagated back to the position $z$. For $z <0$ and $ z^{\prime} > 0$, the Green function is written as
\begin{equation}
G_{yy}(z,z^{\prime}) = \frac{1}{2iq} e^{-iqz}(1+ r_{s}^{0})e^{ikz^{\prime}}~.
\end{equation}
In this case, the Green function the wave propagate from the interface which is proportional to the left-moving wave from $z^{\prime}$ and 
previously reached wave which now reflected back to $z^{\prime}$.

Similar to the argument for s-polarized light, we can write differential equations for the Green  function in p-polarized light case where electric field is in $xz$ plane.
The differential equation for the Green function becomes a coupled equation written as 
\begin{widetext}
\begin{equation}
\left[
\begin{array}{cc}
 \frac{ \partial^{2} }{\partial z^{2}} + \frac{\omega^{2}}{c^{2}} \epsilon^{0}(z) & -iQ  \frac{ \partial   }{\partial z} \\
 -iQ  \frac{ \partial   }{\partial z} & -Q^{2}+ \frac{\omega^{2}}{c^{2}} \epsilon^{0}(z)
\end{array}\right]
\left[
\begin{array}{cc}
G_{xx}(z,z^{\prime}) & G_{xz}(z,z^{\prime} ) \\
G_{zx}(z,z^{\prime}) & G_{zz}(z,z^{\prime})
\end{array}
\right]
= \left[
\begin{array}{cc}
\delta(z-z^{\prime}) & 0 \\
0 & \delta(z-z^{\prime})
\end{array}\right]~.
\end{equation}
\end{widetext}
The solution of the coupled equation is given as 
\begin{eqnarray}
G_{ij}(z,z^{\prime}) &=&  \alpha_{j} [\Theta(z-z^{\prime}) U_{i}(z)V_{j}(z^{\prime}) \nonumber \\
&&+ \Theta(z^{\prime}-z)V_{i}(z)U_{j}(z^{\prime})] \nonumber \\
&+& \delta_{ij}\delta_{iz} \frac{c^{2}}{\epsilon^{0}(z)\omega^{2}}\delta(z-z^{\prime})~,
\label{eq:GP}
\end{eqnarray}
where
\begin{eqnarray}
&& \alpha_{\left\{\substack{x\\z}\right\}} =\Bigg\{
\begin{array}{cc}
 \frac{1}{4i}\frac{q/\epsilon^{0}_{L} 
+ k/\epsilon^{0}_{S}}{\omega^{2}/c^{2}} = \frac{c^{2}}{2i\omega^{2}} 
\frac{q/\epsilon_{L}^{0}}{1+r_{p}^{0}} 
= \frac{c^{2}}{2i\omega^{2}} \frac{k/\epsilon^{0}_{S}}{1-r_{p}^{0}}   \\
 -\frac{1}{4i}\frac{q/\epsilon^{0}_{L} 
+ k/\epsilon^{0}_{S}}{\omega^{2}/c^{2}} = -\frac{c^{2}}{2i\omega^{2}} 
\frac{q/\epsilon_{L}^{0}}{1+r_{p}^{0}}
= -\frac{c^{2}}{2i\omega^{2}} \frac{k/\epsilon^{0}_{S}}{1-r_{p}^{0}} 
\end{array}  \nonumber \\
\end{eqnarray}
\begin{eqnarray}
&& U_{x}(z) = 
\bigg\{
\begin{array}{cc}
e^{iqz}-r_{p}^{0}e^{-iqz}	&  z<0 \\
(1-r_{p}^{0})e^{ikz}	& z>0
\end{array} \\
&& V_{x}(z) = 
\bigg\{
\begin{array}{cc}
(1+r_{p}^{0})e^{-iqz}	&  z<0 \\
 e^{-ikz}+r_{p}^{0}e^{ikz}	& z>0
\end{array}  \\ 
&& U_{z}(z) = \frac{iQ}{ \epsilon^{0}(z) \omega^{2}/c^{2}-Q^{2}}  \frac{ \partial  U_{x}(z) }{\partial z} \nonumber \\
&& = \frac{-Q}{ \epsilon^{0}(z) \omega^{2}/c^{2}-Q^{2}} \bigg\{
\begin{array}{cc}
q(e^{iqz}+r_{p}^{0}e^{-iqz})	&  z<0 \\
k(1-r_{p}^{0})e^{ikz}	& z>0
\end{array} \nonumber \\ \\
&& V_{z}(z) = \frac{iQ}{ \epsilon^{0}(z) \omega^{2}/c^{2}-Q^{2}}  \frac{ \partial  V_{x}(z) }{\partial z} \nonumber \\
&& = \frac{Q}{ \epsilon^{0}(z) \omega^{2}/c^{2}-Q^{2}} \bigg\{
\begin{array}{cc}
q(1+r_{p}^{0})e^{-iqz}	&  z<0 \\
 k(e^{-ikz}-r_{p}^{0}e^{ikz})	& z>0
\end{array} \nonumber \\ \\
&& r_{p}^{0} = \frac{\epsilon^{0}_{S}q-\epsilon^{0}_{L}k}{\epsilon_{S}q +\epsilon_{L}^{0} k}~.
\end{eqnarray} 
As in the case of s-polarized light, each set of $\{ U_{x}, U_{z}\}$ and $\{V_{x}, V_{z} \}$ is the solution of Eq.~(\ref{eq:E0}), the homogeneous wave equation. Thus, 
the Green function also satisfies the asymptotic boundary condition of electric field ($E_{x}\propto V_{x} \sim  e^{-iqz}, E_{z}\propto V_{z} \sim  (Q/q)e^{-iqz}$) in the limit of $z \rightarrow -\infty$. In Eq.~(\ref{eq:GP}), there is one addition term for $G_{zz}$ components, which is required to give zero for the equation
\begin{equation}
\left[\frac{\partial^{2}}{\partial z^{2}}+ \frac{\omega^{2} \epsilon^{0}(z)}{c^{2}} \right]
G_{xz}(z,z^{\prime}) -iQ \frac{\partial }{\partial z}G_{zz}(z,z^{\prime}) = 0~,
\end{equation}
and 
Similar to the $G_{yy}$ in s-polarized light, we can write the Green function $G_{xx}$ for 
p-polarization case for $z, z^{\prime} > 0$ as 
\begin{equation}
G_{xx}(z,z^{\prime}) = \frac{c^{2}k}{2i \omega^{2} \epsilon^{0}_{S}}
 \left[	e^{ik\vert z-z^{\prime}\vert}+r_{p}^{0}e^{ik(z+z^{\prime})}\right]
\end{equation}
and for $z >0, z<0$ as
\begin{equation}
G_{xx}(z,z^{\prime}) = \frac{c^{2}q}{2i \omega^{2} \epsilon^{0}_{L}} e^{-iqz}
 (1-r_{p}^{0})e^{ikz^{\prime}}~,
\end{equation}
which can be interpreted as the same way as the s-polarization case. 
For $G_{zz}(z-z^{\prime})$ for $z, z^{\prime} >0$, we have 
\begin{eqnarray}
G_{zz}(z,z^{\prime}) &=& \frac{Q^{2}}{2ik}\frac{c^{2}}{\omega^{2}\epsilon^{0}_{S}}
	\left[ e^{ik\vert z-z^{\prime}\vert} - r_{p}^{0} e^{ik(z+z^{\prime})}\right] \nonumber \\
&+& \frac{c^{2}}{\epsilon_{S}^{0}\omega^{2}} \delta(z-z^{\prime})~.
\label{eq:Gzz}
\end{eqnarray}
The delta function emerging in $G_{zz}$ accounts for the discontinuity in the electric field due to the current in the $z$ direction. We can eliminate the discontinuity by introducing the displacement $D_{z}$ that is continuous across the interface. 

\begin{figure}[htbp]
\begin{center}
\includegraphics[width=0.9\columnwidth, angle=-0]{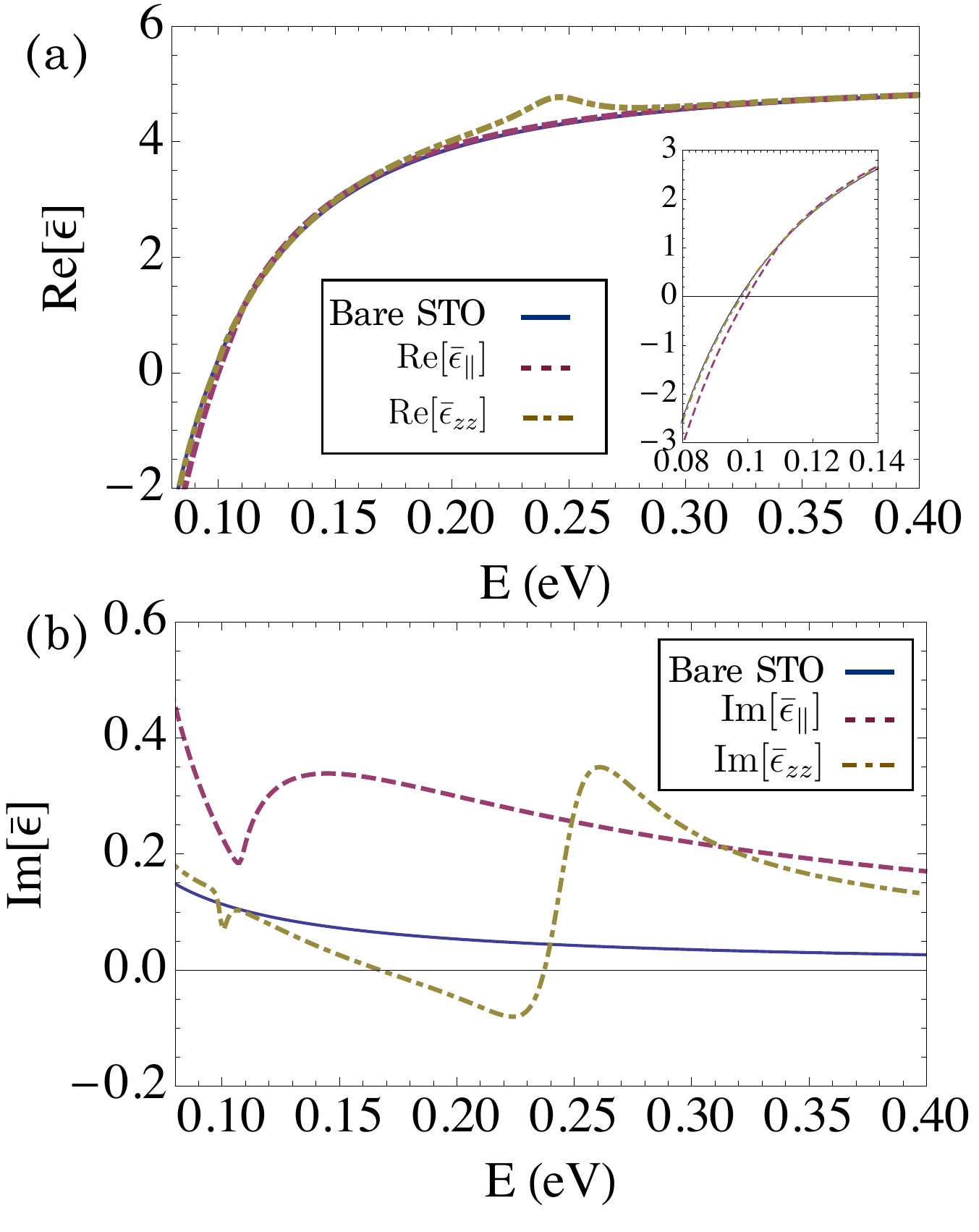}
\caption{(color online) Effective anisotropic dielectric constant for $n_{0} = 0.3$. (a) The real part and (b) imaginary part of anisotropic dielectric tensor  compared with bare STO dielectric constant. Inset: magnification of the low frequency region.}
\label{fig:6}
\end{center}
\end{figure}

\section{Effective anisotropic dielectric tensor}
We introduce an anisotropic dielectric tensor $\bar\epsilon_{\parallel}$ and $\bar\epsilon_{zz}$ as an alternative description of the optical responses. The formulae are obtained by comparing the obtained expression of reflectivity (Eqs.~(\ref{eq:rs-appr}) and (\ref{eq:rp-appr})) to the general expression of reflectivity in anisotropic media.

The reflectivity of $s$- and $p$- polarized light for anisotropic materials with cubic symmetry are given by\cite{Barker73}
\begin{eqnarray}
r_{s} &=& \frac{\sqrt{\epsilon_{L}^{0}} \cos \theta + \sqrt{\bar \epsilon_{\parallel}-\epsilon_{L}^{0} \sin^{2} \theta }}{\sqrt{\epsilon_{L}^{0}} \cos \theta -  \sqrt{\bar \epsilon_{\parallel}-\epsilon_{L}^{0} \sin^{2} \theta }}~, \nonumber \\
r_{p} &=& \frac{\cos \theta / \sqrt{\epsilon_{L}^{0}} - \sqrt{(\bar\epsilon_{zz} - \epsilon_{L}^{0} \sin^{2} \theta)
	/(\bar\epsilon_{zz} \bar\epsilon_{\parallel})}}{\cos \theta / \sqrt{\epsilon_{L}^{0}} + \sqrt{(\bar\epsilon_{zz} - \epsilon_{L}^{0} \sin^{2} \theta)
	/(\bar\epsilon_{zz} \bar\epsilon_{\parallel})}}~,
\label{eq:loc-rp-rs}
\end{eqnarray}
where $\bar \epsilon_{\parallel}(\omega)$ and $\bar \epsilon_{zz}(\omega)$ are anisotropic dielectric tensors for fields parallel and perpendicular to the interface, respectively. By the comparison with  Eqs.~(\ref{eq:rs-appr}) and (\ref{eq:rp-appr}), we obtain 
\begin{widetext}
\begin{eqnarray}
\bar\epsilon_{\parallel}(\omega) &=& \epsilon_{S}^{0} - 2i kd \langle \langle \Delta \epsilon_{\parallel}  \rangle \rangle
	-\frac{\omega^{2} d^{2}}{c^{2}}
	  \langle \langle \Delta \epsilon_{\parallel}  \rangle \rangle^{2} \nonumber \\
\bar \epsilon_{zz}(\omega) &=& \frac{\bar\epsilon_{\parallel} + \frac{\omega^{2} d^{2} \epsilon_{L}^{0}}{c^{2}\epsilon_{S}^{0}}
	 \sin^{2}\theta \langle \langle \Delta \epsilon_{\parallel}  \rangle \rangle^{2} }{1- 2ikd (\epsilon_{S}^{0})^{-1}
	 \left\{\epsilon_{S}^{0}\bar\epsilon_{\parallel}\langle \langle \Delta \epsilon_{zz}^{-1} \rangle \rangle 
	  +  \langle \langle \Delta \epsilon_{\parallel}  \rangle \rangle\right\} +\frac{\omega^{2} d^{2}}{c^{2}}\sin^{2}\theta \epsilon_{S}^{0}\epsilon_{L}^{0}
	  \bar\epsilon_{\parallel}\langle \langle \Delta \epsilon_{zz}^{-1} \rangle \rangle^{2}}~.
\label{eq:bar-rp-rs}
\end{eqnarray}
\end{widetext}
We note that the obtained dielectric constants depend on the incident angle $\theta$. 
We can think of $\bar \epsilon_{\parallel}$ and $\bar \epsilon_{zz}$ as an effective dielectric tensor which generates the same reflectivity as is given by our full calculation. Up to the first order of $\omega d/c$,  Eqs.~(\ref{eq:bar-rp-rs}) take the simple form, 
\begin{eqnarray}
\bar\epsilon_{\parallel} &\simeq& \epsilon_{S}^{0} - 2ikd\langle \langle \Delta \epsilon_{\parallel}  \rangle \rangle \nonumber \\
\bar \epsilon_{zz} &\simeq& \epsilon_{S}^{0} + 2 i kd  (\epsilon_{S}^{0})^{2} \langle \langle \Delta \epsilon_{zz}^{-1} \rangle \rangle~. \nonumber 
\end{eqnarray}

Fig.~\ref{fig:6} shows the real and imaginary part of the effective dielectric tensor for $n_{0}=0.3$ with incident angle $75^{\circ}$ compared with bare dielectric constant of STO. We can see the significant change in both the real and imaginary part contributed from the out-of-plane dielectric tensor around plasmon peak ($\sim 0.25 eV$). The in-plane dielectric tensor mostly contributes to the imaginary part of the dielectric tensor. In addition to the overall increases in the imaginary part of in-plane dielectric constant there is a dip near the $\omega_{pe}$ that is due to the rapid change in $kd$ around LO phonon frequency of STO. The obtained dielectric tensor provides an alternative way to interpret the reflectivity obtained in Sec.~\ref{Model} (c).

\bibliography{park_manuscript.bib}

\end{document}